# Unifying the Temperature Dependent Dynamics of Glasses

Joseph B. Schlenoff[*,1] and Khalil Akkaoui[1]

[1]Department of Chemistry and Biochemistry, The Florida State University, Tallahassee, Florida 32306-4390, United States

## Abstract

Strong changes in bulk properties, such as modulus and viscosity, are observed near the glass transition temperature, $T_g$, of amorphous materials. For more than a century, intense efforts have been made to define a microscopic origin for these macroscopic changes in properties. Using transition state theory, we delve into the atomic/molecular level picture of how microscopic localized relaxations, or "cage rattles," translate to macroscopic structural relaxations above $T_g$. Unit motion is broken down into two populations: (1) simultaneous rearrangement occurs among a critical number of units, $n_\alpha$, which ranges from 1 to 4, allowing a systematic classification of glasses that is compared to fragility; (2) near $T_g$, adjacent units provide additional free volume for rearrangement, not simultaneously, but within the "primitive" lifetime, $\tau_1$, of one unit rattling in its cage. Relaxation maps illustrate how Johari-Goldstein β relaxations stem from the rattle of $n_\alpha$ units. We analyzed a wide variety of glassy materials, and materials with glassy response, using literature data. Our four-parameter equation fits "strong" and "weak" glasses over the entire range of temperatures and also extends to other glassy systems, such as ion-transporting polymers and ferroelectric relaxors. The role of activation entropy in boosting preexponential factors to high "unphysical" apparent frequencies is discussed.

# Introduction

Below their glass transition temperature, $T_g$, amorphous materials exist in a glassy state without long range order. A molecular scale description of the approach to this state from high temperatures, signaled by a rapid increase in bulk stiffness, has long challenged theorists.[1-4] Figure 1 illustrates the dilemma: at sufficiently high temperature, T, the bulk structural relaxation rate, $\omega_\alpha$, mirrored by the viscosity, η, shows classical Arrhenius behavior $ln\omega_\alpha \sim -1/T$, typically used to describe thermally activated processes. Approaching $T_g$, in the "supercooled" region there is a strong deviation from Arrhenius, and at $T_g$ the material becomes a solid glass with almost no liquid-like response.[5] It is generally agreed that cooperative motion between units comprising the glass increasingly limit dynamics approaching $T_g$.[6-7]

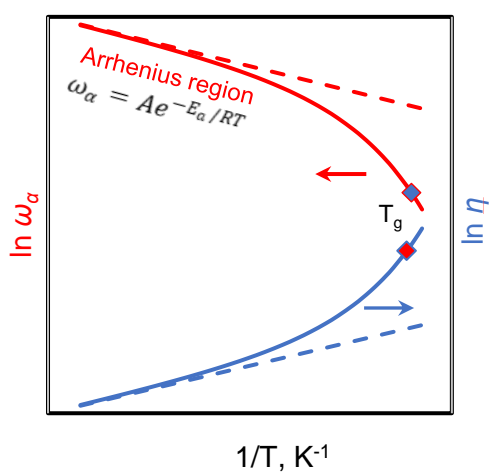

**Figure 1.** Relaxations in glass formers. The dynamics, given by the structural relaxation rate $\omega_\alpha$ (s$^{-1}$), or viscosity $\eta$ (Pa s), slow as it cools. At high temperatures, dynamics can be described by an Arrhenius equation (dashed lines). Deviation from Arrhenius behavior is observed approaching $T_g$ (diamond).

A major goal of theory is to describe dynamics in all glass formers with one equation.[1] Towards this goal, "universal" scalings and relationships have been established. For example, the α-relaxation of many glass formers adheres to thermodynamic scaling: $\omega_\alpha$ is a function of a scaled relationship between density (pressure) and temperature.[8-9] While activation energy $E_a$ and a prefactor A are the only two fit parameters required to describe Arrhenius response, the most economical analytical description of the curvature in the supercooled region needs at least three parameters. Towards this end, the Vogel-Fulcher-Tammann (VFT) equation has been used for several decades;[10]

$$\omega_{T,VFT} = \omega_{0,VFT}\, e^{\frac{-DT_o}{T-T_o}} \qquad [1]$$

where $\omega_{0,VFT}$, D, and $T_o$ are empirical, freely-adjustable fit parameters. $\omega_0$ ($\eta_0$) comes from a high-temperature extrapolation. $T_0$, known as the Vogel temperature, is usually about 50 K below $T_g$. Many decades of discussion and controversy have ensued over the meaning of $T_0$ and whether, as it implies, flow ceases completely at this temperature.[11] VFT fits with one set of parameters are typically useful over a limited temperature range. If the range is expanded, two or more VFT fits,[12-14] or a combination of Arrhenius plus VFT,[15] are often used. Many of the models describing the temperature dependence of structural relaxation have recently been summarized by Novikov and Sokolov.[16]

Eyring modified his transition state theory (TST) to describe the viscosity of liquids[17]

$$\eta = \kappa \eta_0 e^{-\Delta S^*/R} e^{\Delta H^*/RT} \qquad [2]$$

where $E_a = \Delta H^* - T\Delta S^*$, and $\Delta H^*$ and $\Delta S^*$ are enthalpy and entropy of activation, respectively. κ is a transmission coefficient which represent a probability of reaction. Eyring assumed[17] that κ was "probably very nearly unity." Prefactor $\eta_0$ was estimated to be $N_A h/V$ where $N_A$ is Avogadro's number, V the molal volume and h is Planck's constant. In a transition state model, Avramov and Milchev, AM, assumed that dynamics resulted from hopping of units between coordination spheres with a random distribution of activation energies and relaxation time, τ, around a mean.[18] This introduced dynamic heterogeneity and yielded the following three-parameter expression

$$log\eta = log\eta_0 + \left[\frac{\tau}{T}\right]^\alpha \qquad [3]$$

Maxwell's equation relates relaxation rate to viscosity via the high frequency glassy modulus $G_0$: $\omega_0 = G_0/\eta_0$. Extrapolations of AM to high temperatures (i.e. into the Arrhenius region) significantly overestimate $\eta_0$, which should be of order $10^{-2}$ Pa s.[11] Hrma et. al recently concluded[19] that any analytical model representing a full 11-12 orders of dynamic range in viscosity or relaxation time, from $T_g$ up, must have at least four adjustable parameters.

Macedo and Litovitz[20] focused on the deviation from Arrhenius in the supercooled region by breaking down the jump probability of a molecule to an empty site, $p_j$, into the probability of having sufficient energy to break bonds $p_E$ (which is the $e^{-\frac{E_a}{RT}}$ term) and the probability that there is sufficient local free volume for a jump to occur, $p_v$

$$p_j = p_E \times p_v \qquad [4]$$

For the $p_v$ term they employed the expression of Cohen and Turnbull[21] for the probability of finding the minimum local free volume for a jump

$$p_v = e^{-\gamma v_0/v_f} \qquad [5]$$

Where γ is an overlap correction between 0.5 and 1, $v_0$ is the close-packed molecular volume and $v_f$ is the free volume. This yields the following

$$\eta = \eta_0 e^{\gamma v_0/v_f} e^{E_a/RT} \qquad [6]$$

This "hybrid" 5-term equation was able to fit both the supercooled and Arrhenius region of the glasses tested.

An expression by Mauro et al.[22], previously proposed empirically by Waterton,[23] was derived starting from the Adam-Gibbs (AG) equation relating the configurational entropy, $S_c$, of a glass former to the viscosity[24]

$$ln\eta = ln\eta_0 + \frac{E_a s_c^*}{RTS_c} \qquad [7]$$

where $s_c^*$ is a constant high temperature configurational entropy (found in the Arrhenius region) and $S_c$ decreases into the supercooled region. Calculating[25-26] or measuring[27] $S_c$ is not straightforward. Mauro et al. used a temperature-dependent constraint model to obtain $S_c$, assumed the glass transition was defined by a shear viscosity of $10^{12}$ Pa s, and used experimental slopes of log$\eta$ vs. 1/T at $T_g$ (known as "fragilities" [28]) to arrive at an equation of the form

$$log\eta = log\eta_0 + \frac{K}{T}e^{C/T} \qquad [8]$$

Where K and C are constants. A comparison with VFT and AM models concluded that Eq. 8 provided a more accurate description.[22] However, all the above three-parameter equations fail to some extent into the Arrhenius region.[19]

Adam and Gibbs[24] recognized the central importance of correlating molecular (microscopic) motion to structural (bulk) relaxation in glass formers when they ascribed the temperature dependent relaxation rates, $\omega_{\alpha,AG}$ to rearrangements in microstructures termed cooperatively rearranging regions (CRRs) of minimum size $n_T$ units (which they called z*)[24]

$$\omega_{\alpha,AG} = Ae^{-n_T E_1/RT} \qquad [9]$$

where $E_1$ is the activation energy for one unit. At temperatures closer to the glass transition, the apparent activation energy $n_T E_1$ (see the slopes in Figure 2) increases substantially, becoming larger than the enthalpy of vaporization by almost a factor of 5.[24, 29] The increase in apparent $E_a$ is interpreted to stem from increased cooperativity between units.[24]

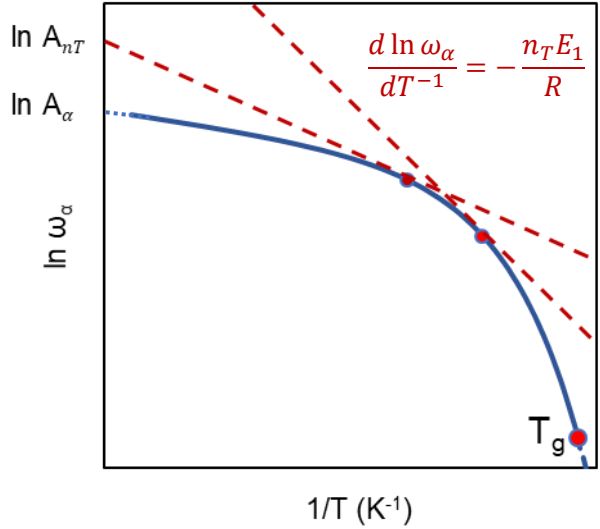

**Figure 2.** $\ln \omega_\alpha$ (s$^{-1}$) as a function of 1/T; solid blue curve. Near the glass transition, $n_T$ cooperatively rearranging units present an increasing *apparent* energy barrier of height $n_T E_1$ (dashed red lines), where $E_1$ is the activation energy for one unit. The prefactor $A_{n(T)}$ also increases strongly with $n_T$ on these Arrhenius plots.

The configurational entropy-based approach in Equation 7 is appealing because the activation energy and $\log\eta_0$ terms remain constant, leaving the configurational entropy to be determined. In contrast, in the AG dynamic cooperativity framework (Equation 9) each of the Arrhenius-style plots (Figure 2) has slope $n_T E_1$ but also a different A intercept which depends on $n_T$. These values of A can be enormous and complicate the original AG approach, which assumed a constant preexponential factor.[24]

In the present work, we re-evaluate the CRR aspects of the AG microscopic model in a transition state theory framework and add a crucial refinement to the idea that local motion on the microscopic scale is described as "caged" motion, where units "rattle" locally to the limits of their cage, but not beyond.[4, 30] Under the right conditions, units escape their cages and rearrange - a structural change impacting bulk properties (e.g. viscosity and modulus). In our analysis, we break cooperative motion down into two types: a small number of correlated units rearranges *simultaneously*, facilitated in the supercooled region by neighbors that move *within the relaxation time of one unit cage rattle*. Separating the microscopic components in this way allows us to separate the *spatial* and *temporal* elements of the α-relaxation and show how localized caged motion evolves into structural relaxation.

## Methods and Model

*Single Units*. At the microscopic level a material consists of single units of minimum size, often presented as beads in simulations.[3, 30-31] At all temperatures above zero, each unit oscillates from its average local position at an *average* (relaxation) rate $\omega_1$

$$\omega_1 = A_1 e^{-E_1/RT} \qquad [10]$$

with activation energy $E_1$ and preexponential factor $A_1$. This frequency refers to units rattling to the limit of their transient cages[4] with an *average* temperature dependent relaxation time $\tau_1 = 1/\omega_1$. In other words, a fraction of units, given by the Boltzmann distribution, $e^{-E_1/RT}$, has enough energy to move to the limit of their cage. The probability that a *specific* unit will move in its cage during time interval $\tau_1$ is 1/e from Poisson statistics (see Section S1). Therefore, the probability that *n(T) specific* neighboring units will move in interval $\tau_1$ is $e^{-n(T)}$.

*The Critical Relaxation Cluster, $n_\alpha$*. Dielectric, mechanical, and thermal methods reveal many relaxation modes in glasses. Johari and Goldstein drew attention to certain relaxations in rigid glasses, $\beta_{JG}$, which had the same Arrhenius temperature dependence as the liquid state (at far higher temperatures) and suggested this relaxation corresponded to less cooperative but caged motion of an unspecified number of adjacent or clustered units.[32-33] Ngai and coworkers have stressed the importance of these "primitive relaxations" as precursors to the α-relaxation at $T > T_g$.[8, 34] Recent computer simulations have supported the evolution of caged dynamics from far below $T_g$ to unhindered dynamics in the liquid state far above $T_g$.[35]

Here, we assume for α-relaxation a cluster of a specific number of units, $n_\alpha$, must rearrange *simultaneously* such that the total energy is the same before and after the move. These $n_\alpha$ units, in a local minimum energy environment, are connected by chemical (covalent, ionic, or metallic bonding) or physical (dipolar, hydrogen bonding) interactions, depending on the material, and move to conserve the symmetry and energy of interaction. This (temperature independent) critical number of units is often assumed to be equal to 1,[24] but, as shown here, it can range from 1 to 4. Above $T_g$, $n_\alpha$ units start to rearrange, below $T_g$, they mostly rattle (see Scheme 1 for an example of $n_\alpha = 2$). If $n_\alpha$ is the same for α and $\beta_{JG}$ relaxations, the activation energy for $n_\alpha$ units rattling *simultaneously* is $n_\alpha E_1 = E_a$, which yields the corresponding Arrhenius form of the *caged* relaxation rate $\omega_{n_\alpha, rattle} = \omega_{JG} = A_{JG} e^{-E_a/RT}$, where $A_{JG}$ is the prefactor for rattling (see Figure S1, Supplemental Information).

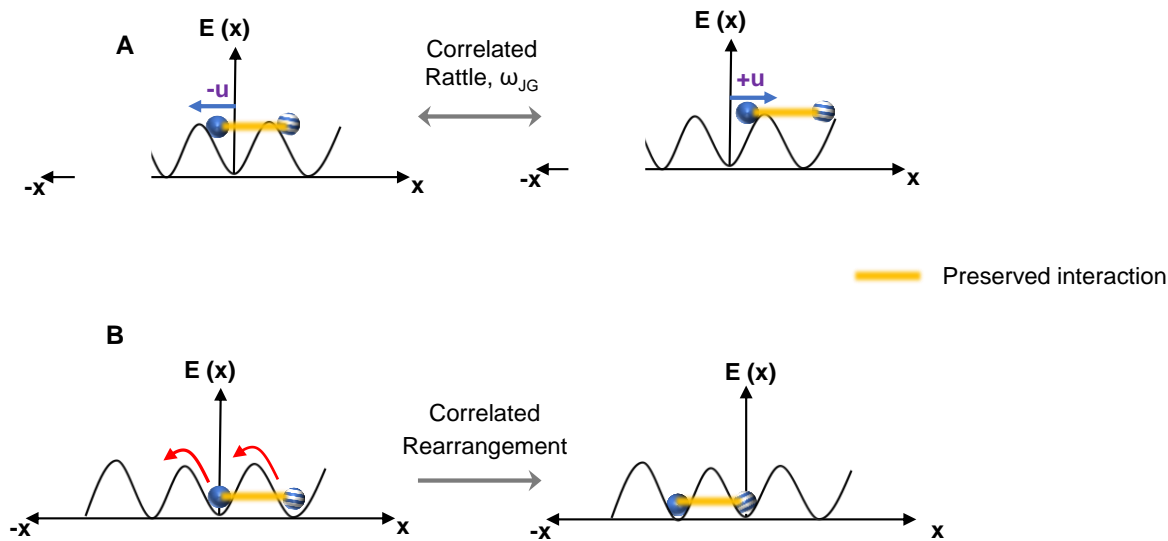

**Scheme 1. A.** A rattle of two units, which move in a correlated manner to preserve the interaction between them. **B.** Some rattles evolve into a rearrangement, seen as a structural relaxation.

*Cooperatively Assisting Units, $n_c$.* When the amplitude of the rattle is enough to overcome the boundaries of the cage, a unit, or a cluster $n_\alpha$ of them, may escape. When does a caged relaxation evolve into a structural relaxation? Doolittle and others advanced the idea that the degree of structural mobility is connected directly to the free volume of the material[36-37] and cooperativity is a consequence of limited free volume.[7, 38] Thus, $n_\alpha$ units require at least $V_\alpha$ of free volume to rearrange i.e. to go from caged to structural relaxation (see Section S2). At sufficiently high temperatures (the Arrhenius region) there are no (free volume) constraints to the concerted movement of $n_\alpha$ units. α-relaxation occurs with the same slope as caged $n_\alpha$ but the intercept is higher because $A_\alpha > A_{JG}$. As the temperature decreases, free volume also decreases. At some specific temperature, $T_{SA}$ (SA = super-Arrhenius), at least one of the neighboring units must move to create sufficient free volume during interval $\tau_1$, which occurs with probability $1/e$. This additional unit, and other neighbors that are increasingly called upon as the temperature decreases, are termed "cooperatively assisting," $n_{c(T)}$. Their purpose is to focus the local free volume, which fluctuates spatially and temporally throughout the sample, round $n_\alpha$. Thus, the minimum total number of neighboring units moving cooperatively within $\tau_1$ at any temperature, $n_T = n_\alpha + n_{c(T)}$.

$T_{SA}$ marks the start of deviation from Arrhenius response of the α-relaxation where it now enters the spatio-temporally restricted regime. Microstructures of $n_\alpha$ units moving while conserving the symmetry of interaction are suggested in Figure 4A - C. From the results below, "simple" molecular glasses with mainly van der Waals intermolecular interactions (no hydrogen bonding) appear to yield $n_\alpha = 1$. Polymers appear to be a special case wherein $n_\alpha$ is always $= 2$. For obvious reasons, movement in linear polymers is dimensionally constrained due to the connectivity of the repeat units: only a small segment of the chain, such as the persistence length, can move as a unit. Figure 4D shows the exchange of two polymer segments as a possible example of a $n_\alpha = 2$ concerted move. Because the Si atom in silicates shares oxygens with four other Si, the $n_\alpha = 4$ found for silicates and related glasses could mean that a "unit" is an oxygen atom. Using high-temperature $^{29}$Si NMR measurements of $K_2Si_4O_9$, relaxation consistent with exchange of oxygens was identified with a total activation energy $E_a$ of 180 kJ mol$^{-1}$,[39] in the range of $E_a$ for silicates listed below.

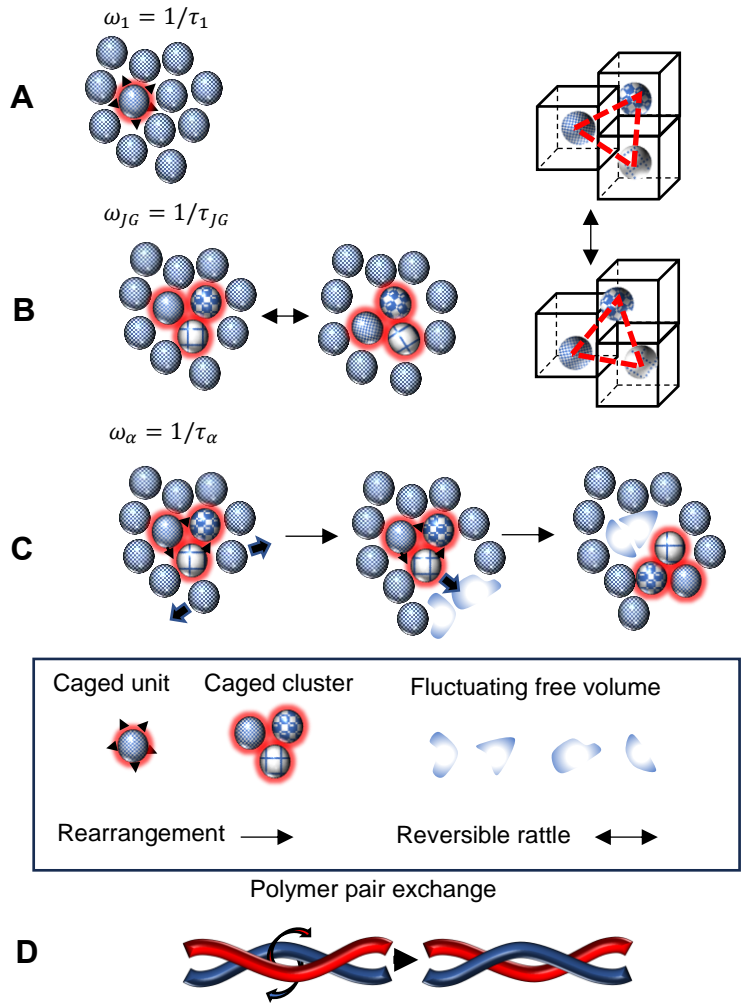

**Figure 4.** Relaxation modes. **A & B** caged motion of 1 or 3 units, the latter sampling the extremes of the cage while maintaining its $n_\alpha = 3$ interaction environment; **C)** $\tau_\alpha$: structural relaxation of $n_\alpha = 3$ units facilitated by two neighbors ($n_{c(T)} = 2$) which provide enough free volume for $n_\alpha$ units to structurally rearrange. **D**. Possible "unit" exchange in polymers.

There are no preordained $n_\alpha$ clusters. Within any $\tau_1$, any cluster of $n_\alpha$ units may attempt to rearrange. Below $T_{SA}$ they are not assured of the free volume needed to rearrange, so they have to rely on $n_{c(T)}$ neighbors to move during $\tau_1$ to provide the required free volume.

***Heterogenous Dynamics.*** Another generally accepted feature of supercooled liquids is that they exhibit spatially heterogenous dynamics,[1-3] or regions of different sizes moving at different rates.[40] In the present analysis, we incorporate the central idea of dynamic heterogeneity as follows: first, there is a spatially- and temporally- fluctuating distribution of sizes in the CRR. Second, as with AG, at any temperature all CRRs above a minimum size, $n_T$ (= $n_\alpha$ + $n_{c(T)}$), contribute to bulk α-relaxation:

$$\omega_\alpha = A'_\alpha e^{-E_a/RT} \sum_{n=n_{c(T)}}^{\infty} e^{-n_T} \qquad [11]$$

Any CRR of size $n_T$ and above contributes to α-relaxation with weight $e^{-n_{c(T)}}$. The summation is a geometric sum and therefore Equation 11 becomes

$$\omega_\alpha = A'_\alpha e^{-E_a/RT} \frac{e^{1-n_{c(T)}}}{e-1} = A_\alpha e^{-E_a/RT} e^{-n_{c(T)}} \qquad [12]$$

where $A_\alpha = \frac{e}{e-1} A'_\alpha$.

In TST, $E_a$ is broken down into activation entropic and enthalpic contributions to the activation barrier. For the unit (n = 1) *rattle*

$$\omega_1 = A_0 e^{\Delta S_1^*/R} e^{-\Delta H_1^*/RT} \qquad [13]$$

and for the $n_\alpha$ *rattle*

$$\omega_{JG} = A_0 e^{\Delta S_{JG}^*/R} e^{-\Delta H_{n\alpha}^*/RT} \qquad [14]$$

This presentation makes it clear that the intercept of an Arrhenius plot includes a physical attempt frequency, $A_0$, which is assumed to be near the Boson frequency, *and* an activation entropy contribution. Only when ΔS* is negligible does the A intercept represent an attempt frequency. The (overlooked) importance of ΔS* has recently been emphasized by Xu et al.[41] The full TST equation for the structural relaxation is

$$\omega_\alpha = A_0 e^{\Delta S_{struct}^*/R} e^{-\Delta H_{n\alpha}^*/RT} e^{-n_{c(T)}} \qquad [15]$$

where $\Delta S_{struct}^*$ is the structural activation entropy for $n_\alpha$ units and $\Delta H_{n\alpha}^* = n_\alpha \Delta H_1^*$. The deviation of $\omega_\alpha$ from Arrhenius at any temperature is simply a factor of $e^{-n_{c(T)}}$.

Equation 12 shows how relaxation depends on a Boltzmann distribution term, $e^{-E_a/RT}$, representing the steady-state fraction of $n_\alpha$ clusters with energy sufficient to overcome barrier $E_a$, and a statistical term, $e^{-n_{c(T)}}$, stemming from the limiting of free volume with decreasing temperature below $T_{SA}$ (free volume continues to increase above $T_{SA}$ but no longer limits the dynamics). A final assumption from Adam and Gibbs[24] allows for the substitution of $n_{c(T)}$: the slope of the $\omega_\alpha$ curve is (see Figure 2)

$$\frac{d \ln\omega_\alpha}{d T^{-1}} = -\frac{n_T E_1}{R} \qquad [16]$$

$n_T E_1$ only *appears* to be a growing activation barrier. It represents the slope of an Arrhenius equation (as in Figure 2) for $n_T$ units rearranging *simultaneously*. In our treatment, only $n_\alpha$ units actually rearrange simultaneously. If $n_T E_1$ were the actual barrier, the probability of rearranging at most temperatures would vanishingly small approaching $T_g$. For example, if $A_\alpha$ were $10^{13}$ s$^{-1}$, $E_1$ = 20 kJ, $T_g$ = 300 K, $n_{Tg}$ = 13, then $n_T E_1$ would be 260 kJ and $\omega_\alpha$ would be about $10^{-21}$ s$^{-1}$ instead of approximately 0.01 s$^{-1}$, as usually seen.

At temperatures greater than $T_{SA}$, the free volume around $n_\alpha$ is enough to allow concerted rearrangement without any contributions from $n_{c(T)}$ units ($n_T \to n_\alpha$; $n_{c(T)} \to 0$) and the frequency attains Arrhenius behavior

$$\omega_{\alpha,Arr} = A_\alpha e^{\frac{-E_a}{RT}} \qquad [17]$$

$A_\alpha$ is extrapolated from higher temperature measurements where $T > T_{SA}$ as shown in Figure 2 and Figure S1. To minimize the error in measured $E_a$ and $A_\alpha$ the temperature should extend as far into the Arrhenius region as possible.

The *deviation* of the measured frequency from the Arrhenius frequency (i.e the frequency calculated at any temperature using Equation 12) is

$$\ln\left(\frac{\omega_{\alpha,Arr}}{\omega_\alpha}\right) = n_T - n_\alpha = n_{c(T)} \qquad [18]$$

The number of cooperative units at $T_g$, $n_{Tg}$ is given by

$$n_{Tg} = \ln\frac{\omega_{\alpha,Arr}}{\omega_{\alpha,Tg}} + n_\alpha \qquad [19]$$

The slope of the linear Arrhenius plot is

$$-\frac{d\ln\omega_{\alpha,Arr}}{dT^{-1}} = \frac{n_\alpha E_1}{R} \qquad [20]$$

where $n_\alpha E_1 = E_a$.

Combining Equations 16 and 18 gives a first order differential equation:

$$\frac{d \ln \omega_\alpha}{d\, T^{-1}} - \frac{E_1}{R} \ln \omega_\alpha + \frac{E_1}{R}\left(n_\alpha + \ln \omega_{\alpha,Arr}\right) = 0 \qquad [21]$$

Expanding the Arrhenius term using Equation 17:

$$\frac{d \ln \omega_\alpha}{d\, T^{-1}} - \frac{E_1}{R} \ln \omega_\alpha + \frac{E_1}{R}\left(n_\alpha + \ln A_\alpha - \frac{n_\alpha E_1}{RT}\right) = 0 \qquad [22]$$

Making the appropriate substitutions yields:

$$\ln \omega_\alpha = \ln A_\alpha - \frac{n_\alpha E_1}{RT} + c_1 e^{\frac{E_1}{RT}} \qquad [23]$$

As $E_a = n_\alpha E_1$, using Equation 17, $\ln A_\alpha - \frac{n_\alpha E_1}{RT} = \ln \omega_{\alpha,Arr}$ which is substituted into Equation 23:

$$\ln \omega_\alpha = \ln \omega_{\alpha,Arr} + c_1 e^{\frac{E_1}{RT}} \qquad [24]$$

To solve for $c_1$, a temperature, $T_{SA}$, where the number of units $n_T$ increases from $n_\alpha$ to $n_\alpha +1$ is defined. It describes the divergence of $\omega_\alpha$ from $\omega_{\alpha,Arr}$ by a factor of $1/e$:

$$\ln\left(\frac{\omega_{\alpha,Arr}(T_{SA})}{\omega_\alpha(T_{SA})}\right) = -c_1 e^{\frac{E_1}{RT_{SA}}} = 1 \qquad [25]$$

This allows us to write $c_1$ as

$$c_1 = -e^{\frac{E_1}{RT_{SA}}} \qquad [26]$$

and substituting $E_1$ with $E_a/n_\alpha$ finally gives

$$\ln \omega_\alpha = \ln \omega_{\alpha,Arr} - exp\left(\frac{E_a}{n_\alpha R}\left(\frac{1}{T} - \frac{1}{T_{SA}}\right)\right) \qquad [27]$$

Assuming the temperature range reaches well into the Arrhenius region (reflected by a linear $\ln \omega_\alpha$ *versus* $1/T$ plot) the only parameter which is not directly and uniquely obtained from the data is $n_\alpha$, which remains the only freely adjustable fit parameter, albeit limited to integer values between 1 and 4 (Table 1). We have presented a form of Equation 27 for polymers but, despite excellent agreement with experimental results, we were not able to rationalize each parameter.[42]

## Results

We tested Equation 27 against experimental data from a wide range of glass formers representing the glassy "universe." We used only data which spanned both the high temperature and supercooled regions, giving $A_\alpha$, $E_a$, and $T_{SA}$ experimentally. The impressive agreement is illustrated in Figure 5, with detail on 16 glasses provided in Table 1 (see Figure S2 for individual plots). Classification of glasses is made according to $n_\alpha$ in Table 1, which also lists the measured $A_\alpha$, $E_a$ and $T_{SA}$. We emphasize that $T_{SA}$ uniquely occurs at $n_T = n_\alpha + 1$ and was also taken directly from the data (see Figures S3 and S4 for detail on how $E_a$ and $T_{SA}$ were determined). Selection of $n_\alpha$ is not arbitrary; Figure S5 shows an example outcome if $n_\alpha$ is selected to be ± 1 of its optimal value (Table 1) or if $T_{SA}$ is off by a few degrees. In contrast, reasonable VFT fits may be obtained with widely different combinations of $\omega_{0,VFT}$, D and $T_0$.[43]

Because $T_g$ in polymers depends on chain length below a certain molecular weight[44] (ca. $10^4$ Daltons), the values of $E_a$ and $T_{SA}$ also depend somewhat on molecular weight. Unlike polymers, alcohols and metallic glass formers cannot be universally classified under one class as shown in Table 1.

**Table 1**. Glass formers whose dynamics are reported in Figure 5 and Figure S2 along with their corresponding $n_\alpha$, glass transition temperature ($T_g$), activation energy ($E_a$), structural preexponential factor $A_\alpha$, fragility m, $T_{SA}$ and the corresponding references.

| Glass Former | $n_\alpha$ | $T_g$ (K) | $E_a$ (kJ mol$^{-1}$) | m | n at $T_g$[a] | $T_{SA}$ (K) | $A_\alpha$ (s$^{-1}$) |
|---|---|---|---|---|---|---|---|
| o-terphenyl (OTP)[29] | 1 | 240 | 23.7 | 81[28] | 26 (16) | 335 | 4.71 x 10$^{12}$ |
| Propylene Carbonate (PC)[15] | 1 | 157 | 14.3 | 132[28] | 23 (28) | 219 | 1.05 x 10$^{12}$ |
| Polystyrene (PS)[42] | 2 | 373 | 106 | 105[42] | 14 (14) | 425 | 3.39 x 10$^{18}$ |
| Poly(vinyl acetate) (PVA)[42] | 2 | 313 | 70.6 | 75.8[42] | 13 (13) | 390 | 1.45 x 10$^{16}$ |
| Poly(isobutyl methacrylate) (PiBM)[42] | 2 | 328 | 61.6 | 64.5[42] | 13 (13) | 421 | 2.83 x 10$^{14}$ |
| Poly(propylene glycol) (PPG)[45] | 2 | 206 | 47.9 | 117[28] | 13 (19) | 250 | 4.55 x 10$^{16}$ |
| Xylitol[46] | 2 | 248 | 63.6 | 94[46] | 20 (14) | 311 | 5.25 x 10$^{18}$ |
| 1-methyl-3-butyl-imidazolium[47] | 2 | 193 | 30.3 | | 19 | 274 | 1.61 x 10$^{13}$ |
| Ni$_{62.4}$Nb$_{37.6}$[48] | 2 | 945 | 165 | 121[48] | 26 (27) | 1353 | 5.34 x 10$^{17}$ |
| Ethanol[49] | 2 | 97 | 17.5 | 61[49] | 15 (13) | 129 | 4.80 x 10$^{12}$ |
| **Pt$_{60}$Ni$_{15}$P$_{25}$**[48] | 3 | 464 | 67.1 | 67[48] | 25 (27) | 990 | 1.14 x 10$^{16}$ |
| **Propanol**[50] | 4 | 103 | 24.1 | 40[28] | 12 (13) | 150 | 8.18 x 10$^{12}$ |
| **Pd$_{40}$Ni$_{40}$P$_{20}$**[48] | 4 | 554 | 104 | 48[48] | 18 (20) | 1053 | 7.85 x 10$^{15}$ |
| **Sodium Germanate**[51] | 4 | 801 | 250 | 46[51] | 14 (11) | 1100 | 4.06 x 10$^{17}$ |
| **Westminster Abbey glass**[52] | 4 | 592 | 164 | 42[52] | 13 (12) | 810 | 1.56 x 10$^{15}$ |
| **Potassium Silicate**[51] | 4 | 739 | 177 | 33[51] | 13 (11) | 1150 | 1.67 x 10$^{13}$ |

$^a n_{T_g}$ is calculated using Equation 19 at T = T$_g$ and the value in parentheses is estimated using $m = \frac{n_{T_g} E_a}{2.3 n_\alpha R T_g}$ and literature values for *m* (superscripts are references).

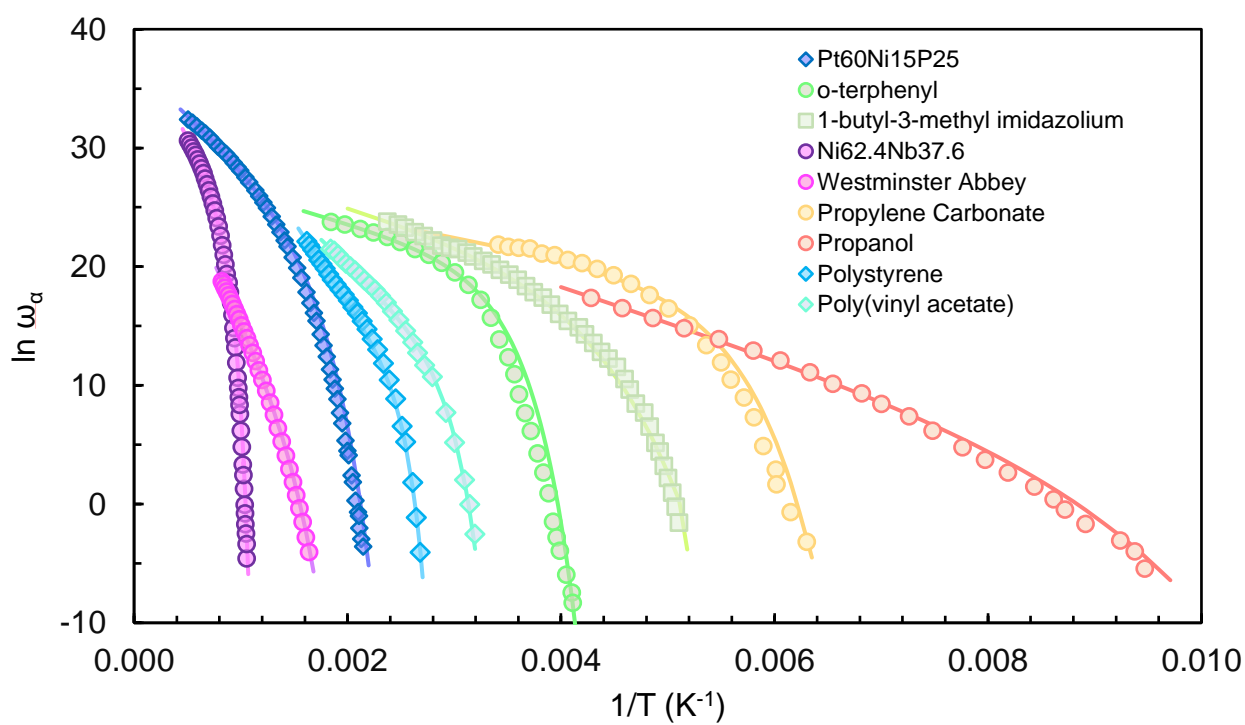

**Figure 5**. Comparison of Equation 27 to experimental data for different types of glasses as a function of 1/T. The α-relaxation rate, $\omega_\alpha$ (s$^{-1}$) of a selection of inorganic, metallic, polymer and small molecule glass formers. Symbols are experimental points from the literature. Solid curves are the results of Equation 27 using the measured parameters listed in Table 1. Only n$_\alpha$ is freely adjustable (but must be an integer between 1 and 4).

## Discussion

Equation 27 relies on a well-accepted TST foundation with the addition of a statistical term $e^{-n_{c(T)}}$, which accounts for deviation from Arrhenius. With the Arrhenius slope, $T_{SA}$ and $n_\alpha$, Equation 27 predicts the entire supercooled region. TST expressions contain a transmission coefficient term, $\kappa$, usually assumed to be 1, but is < 1 if some activated complexes do not give rise to products.[53] Given the probabilistic derivation of $n_{c(T)}$ it can be thought of as a $\kappa$: some groups of $n_\alpha$ units have sufficient energy to rearrange but only a fraction $e^{-n_{c(T)}} = \kappa$ actually do so. Eyring's assumption[17] that $\kappa = 1$ would thus be correct in the Arrhenius regime far above $T_g$ i.e. in the liquid state.

Criteria for whether a secondary relaxation may be considered a "true" $\beta_{JG}$, the precursor to the α-relaxation, have been provided by Ngai and coworkers.[9,8] In broadband dielectric spectroscopy, BDS, for many glasses, $\beta_{JG}$ shows up as a high frequency wing or shoulder on the much more intense α-relaxation.[54] Deconvolution with arbitrary fitting functions may not provide reliable $\beta_{JG}$ values.[55] Thus, the use of glass formers with well-distinguished α- and β-peaks in the BDS is preferred, such as the series of polyols sorbitol, threitol and xylitol reported by Döß et al.[46] Figure 6 shows an example of a relaxation map for xylitol ($n_\alpha = 2$). The primitive $\omega_1$ rattle of one unit, with an Arrhenius slope $-E_1/R$, is the fastest relaxation at all T and extrapolates to $A_1$ at high T. $\omega_{\beta JG}$ with slope $-n_\alpha E_1/R$ describes caged relaxation by $n_\alpha$ units.

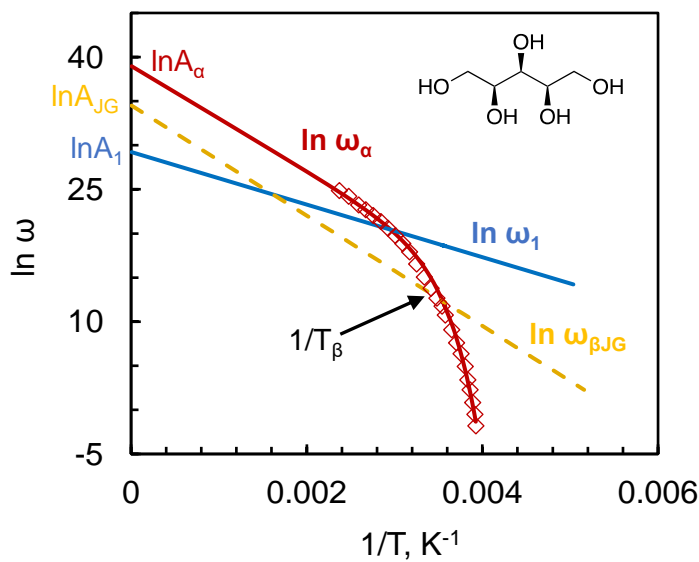

**Figure 6.** Relaxation map for xylitol. Structural relaxation rate $\omega_\alpha$ by cooperative motion among $n_T$ units; open diamonds, experimental from Döß et al.,[46] solid line from Equation 27 with parameters in Table 1. Unit rattle $\omega_1$ is given in blue. $\beta_{JG}$ relaxation of caged clusters of $n_\alpha$ in yellow dash ($\omega_{\beta JG}$) from Döß et al.[46] Corresponding intercepts are $A_{JG}$ for the local, caged rattle; and $A_\alpha$ for the structural relaxation. $A_1$ is the intercept for a single unit rattle assumed to be near the Boson peak (although identifying the actual Boson peak from other modes in the ~1 THz range may not be straightforward[56]).

The α-relaxation in the Arrhenius region differs from the linear extrapolation of $β_{JG}$ by a factor $\frac{e^{\Delta S^*_{struct}}}{e^{\Delta S^*_{JG}}}$ (compare Equations 14 and 15). For glasses having $n_α = 1$, the difference is small and may suggest that $ω_α$ overlaps in the Arrhenius region with an extrapolated $ω_{βJG}$, as was done in Johari's original paper, which extrapolated a β-relaxation near and below $T_g$ to the liquid state.[32] Usually, $ω_{βJG}$ *merges* with $ω_α$ at temperature $T_β$ where $Rn_{c(T=Tβ)} = \Delta S^*_{struct} - \Delta S^*_{JG}$. If $\Delta S_{JG}$* and $\Delta S_{struct}$* are small and $A_0 ≈ A_{JG} ≈ A_α$ then $ω_α ≈ ω_{JG}$. This case is more likely to be satisfied when $n_α = 1$. For example, Angell et al. identified a Boson peak in OTP ($n_α = 1$) at 18 cm$^{-1}$ or 5.4 x 10$^{11}$ Hz x $2π$ = 3.4 x 10$^{12}$ s$^{-1}$ relaxation rate which is close to the $A_α$ value in Table 1.[57] Figure S1 shows the relaxation map for sorbitol and threitol along with experimental data corresponding to their $β_{JG}$ relaxations.

Several words of caution are needed when making the long extrapolation of $lnω_{βJG}$ to the $A_{JG}$ intercept. First, $ω_{βJG}$ is usually obtained over a narrow range of temperatures in the supercooled and below-$T_g$ regions, introducing uncertainty in the slope. The data taken from the literature usually shows a slightly different slope between $lnω_α$ and $lnω_{βJG}$. There may be a change in slope for $ω_{βJG}$ as it passes through $T_g$. However, the $β_{JG}$ data for propanol of Hansen et al. show $E_a$ for $β_{JG}$ = 20 kJ mol$^{-1}$ compared to about 24 kJ mol$^{-1}$ in Table 1. In contrast, the temperature dependence of the relaxation time for propanol measured by Caporaletti et al. using both BDS and nuclear resonance time-domain-interferometry, TDI, had an $E_a$ of about 30 kJ mol$^{-1}$.[58] $β_{JG}$ measured via TDI is better decoupled from the α-relaxation.[55] Whether the material is at equilibrium when $β_{JG}$ is measured has been addressed. Wagner and Richert noted that isothermal aging of OTP, a widely studied GF, near $T_g$ results in the loss of a peak identified as $β_{JG}$ whereas sorbitol does not suffer from this issue.[59]

In general, $A_α$ for $n_α > 1$ is often at apparently unphysical rates (e.g. 10$^{16}$ - 10$^{19}$ s$^{-1}$ in Table 1). Even among polymers, there is great variability in $A_α$, while $ω_{Boson}$ for polymers is clustered around 0.5 THz.[60] Equations 13 – 15 show the role of activation entropy in TST, whatever the process, be it a unit rattle, an $n_α$ rattle, or an $n_α$ structural relaxation. $\Delta S_1$* and $\Delta S_{JG}$* are likely to be < $\Delta S_{struct}$*. Qualitatively, for $n_α > 1$, a *rearrangement* of $n_α$ units involves a greater volume than a *rattle*. Since $n_{C(T)} \rightarrow 0$ at high temperature, the difference $lnω_α - lnω_{JG} = (\Delta S^*_{struct} - \Delta S^*_{JG})/R$. This explains the parallel nature of the $lnω_α$ and extrapolated $lnω_{JG}$ plots at high temperature (Figure 6).

Lawson, in analyzing activated processes in solids, derived an approximate equation for relating the activation entropy to the volume change, $\Delta V$*, of an activated process[61]

$$\Delta S^* = \frac{\delta}{\theta}\Delta V^* \quad [28]$$

Where δ is the volume coefficient of thermal expansion and θ the isothermal compressibility. Polystyrene has the largest $A_α$ of the polymers in Table 1, probably a result of the inherent stiffness (long persistence length) of the backbone. From Eq. 15, $lnA_α - lnA_0 = \Delta S^*_{struct}/R$. For polystyrene, using values of 6.1 x $10^{-4}$ $K^{-1}$ and 12.1 x $10^{-10}$ $Pa^{-1}$ at 300 ºC (well into the Arrhenius regime) for δ and θ, respectively,[62] and $A_0$ = 2.7 x $10^{12}$ $s^{-1}$,[63] $\Delta S^*_{struct}$ is 117 J $mol^{-1}$ $K^{-1}$ and $\Delta V^*$ is 230 $cm^3$ which is 2.2 molar volumes of styrene repeat units, or about one styrene from each of the $n_\alpha$ = 2 units rearranging in the Arrhenius region. Assuming this is the minimum free volume needed to rearrange, supplied at $T_g$ by $n_{Tg}$ = 13 units of persistence length 4 styrene units each, the fractional free volume, $v_f$, at $T_g$, is about 2.2/(13 x 4) = 4 %. Depending on the model, estimates of $v_f$ for polystyrene at $T_g$ range from about 2.5 % (Doolittle model[7,36-37]) to 11 % (White and Lipson using a PVT model[34]) to 12 % (from positron lifetime spectroscopy[64]). Eyring's theory is also used to described plastic deformation in glasses.[17][65] Long et al.[65] modified classical TST theory to account for a mismatch between apparent activation volumes between the linear and plastic flow regimes. In their analysis, elastic energy stored over a length scale of dynamic heterogeneities lowers the energy barrier for α-relaxation.

Relaxation plots in GFs are known to exhibit a thermodynamic scaling[8,9] between density, ρ (controlled by pressure), and temperature: $\omega_\alpha \sim T/\rho^\gamma$. In select cases, for small van der Waals GFs such as OTP and PC, the entire relaxation spectrum in BDS follows this scaling.[8] Such isochronal superposition (IS) could only occur if $e^{\Delta S^*_{struct}} \approx e^{\Delta S^*_{JG}}$ i.e. the activation volumes of $\beta_{JG}$ and $\omega_\alpha$ are similar (according to Equation 28). In addition, the $\omega_\alpha$ broad band spectra at different temperatures superimpose with frequency shifting (time-temperature superposition), e.g. for OTP.[66] These small, van der Waals GFs are likely to have $n_\alpha$ = 1. Niss and Hecksher[66] point out that polymers ironically do not show TTS over the entire BDS response. This is because $n_\alpha$ for polymers = 2.

**Classes of glasses**. Classification of glasses is typically done by comparing the extent to which their dynamics deviate from high-temperature Arrhenius using a steepness index or kinetic fragility, *m*, at $T_g$,[28]

$$m = -\left|dlog_{10}\omega_\alpha/d(T_g/T)\right|_{T=T_g} \quad [29]$$

Using Eqs. 16 and 19, the fragility and $n_\alpha$ are related by

$$m = \frac{n_{T_g}E_a}{2.3n_\alpha RT_g} = \frac{n_{T_g}E_1}{2.3RT_g} \quad [30]$$

which provides a quick estimate of the $n_\alpha$ class using literature $T_g$, *m* and $E_a$ without fitting the data over all temperatures. The origin of fragility is a perplexing issue in glass physics. From Equation 30, the origin of fragility and the reasons for the wide range of values of *m* are clear: it is a combination of three parameters. $n_{Tg}$ has a lower spread of values (about a factor of two) than does $T_g$ or $E_1$. Some of the glass formers with a large $E_a$ actually have much lower $E_1$ if $n_\alpha$ is 4. Thus, the silicate glasses, with $n_\alpha = 4$ and high $T_g$s, push fragility lower. Small molecules with low $T_g$ and van der Waals interactions (where $n_\alpha = 1$) tend toward the upper half of fragilities.

Table 1 shows reasonable agreement between $n_{Tg}$ (n at T = $T_g$) estimated with Equation 30 using the literature *m* values and those calculated using Equation 19. The actual $T_g$ is not predicted from the equations herein, although it can be predicted with the locally correlated lattice (LCL)[21] model, which is a first-principles thermodynamic treatment of PVT data. The LCL model frees itself of the constraint arbitrarily placed on Doolittle's free volume to make it follow VFT response. [21]

Supercooled liquids approach icosahedral ordering[67] near $T_g$ and atomistic models show such ordering improves with observation time.[68] In fact, the CRR is only compact at $T_g$ and, if quasi-icosahedral, would imply $n_{Tg}$ of about 13, representing a central unit surrounded by 12 nearest neighbors. The configuron percolation model[69] supports conclusions by Wendt and Abraham[70] that packing saturation of the first shell by nearest neighbors at $T_g$ is predicted by analysis of the pair distribution function (PDF). Example X-ray scattering studies on PDFs of metallic glasses show a coordination number of about 12.3[71] and 12.6[72] From Table 1, many $n_{Tg}$ are clustered at around 13 but this is not universal, probably reflecting anisotropy[73] in the interaction environment of a unit, such as the hydrogen bonding patterns in xylitol and glycerol. Note that polymers are structurally anisotropic (they are chains) but each moving unit might experience the same interaction environment.[42] $n_{Tg}$ is sensitive to the slope of the fit at $T_g$, which is least satisfactory for the OTP fit. The measured slope from the OTP data near $T_g$ (see Figure S2) gives $n_{Tg} = 13$. Overall, the answer to "what happens at $T_g$?" is, for many systems, "that is the lowest temperature where 13 neighboring units with a near-icosahedral interaction environment can cooperatively rearrange within $\tau_1$ to provide the free volume that allows $n_\alpha$ units to rearrange simultaneously." $n_{Tg}$ may be larger or smaller if the environment does not have icosahedral symmetry, or the units themselves adopt different ordering as required, for example, by hydrogen bonding. Dynamic modeling that starts with spherically symmetric units (e.g. beads) would be expected to achieve idealized short-range ordering approaching close packing.

If cooperativity in the first coordination shell is "saturated" at $T_g$, what happens when the glass is further cooled? The next-nearest neighbor shell must participate. Modeling of amorphous materials by Mercier and Levy[74] suggests that the packing goes from 13 units to 43 if the next coordination shell is included (for spherical units). One can expect another glass transition at a lower temperature where the entire shell of next-nearest neighbors must also cooperatively rearrange. The changes between coordination shells probably underly second-order like transitions such as a change in expansion coefficient and a step in heat capacity at $T_g$. Equation 27 is not extended below $T_g$ as there is no basis to assume $E_a$ remains constant below $T_g$. In particular, the activation energy term in Equation 15 may change. There is evidence that below $T_g$ the relaxation response is Arrhenius with a much higher $E_a$ than that above $T_g$.[75]

The physical size of $n_{Tg}$ units is reasonably consistent with experiment. For example if the actual "unit" in polystyrene is a persistence length of 1 nm which is 1/0.25 = 4 styrene repeat units and there are 13 of these at $T_g$, the mass of the CRR is 4 x 13 x 104 g mol$^{-1}$/6.02 x 10$^{23}$ mol$^{-1}$ = 9.0 x 10$^{-21}$ g or 9.0 x 10$^{-20}$ cm$^3$ assuming a density of 1 g cm$^{-3}$ for a size of 2.1 nm, compared to 3.0 nm estimated by Hempel et al.[76]

**Shift Factors.** Equation 27 can be rearranged into classical shift factors using the parameters $E_a$, $T_{SA}$ and $n_\alpha$ (Section S4). In the broadly-used concept of time-temperature superposition, TTS, a shift factor $a_T$ is employed to shift dynamic responses taken at different temperatures and is the ratio between their frequencies, $\frac{\omega_{T_1}}{\omega_{T_2}}$. A typical expression for shift factors in polymers is that of Williams, Landel and Ferry (WLF):

$$\log(a_T) = \frac{-c_1(T_1-T_2)}{c_2+T_1-T_2} \qquad [31]$$

where $c_1$ and $c_2$ are two empirical (freely adjustable) fit constants.[77] Using the appropriate substitutions (Section S4) and Equation 16, we obtain the following:

$$\ln a_T = \frac{E_a}{R}\left(\frac{1}{T_2}-\frac{1}{T_1}\right) + e^{\frac{-C}{T_{SA}}}\left(e^{\frac{C}{T_2}} - e^{\frac{C}{T_1}}\right) \qquad [32]$$

where $C = E_a/n_\alpha R$. This relationship shows there are two components to the shift factor; the first term on the right represents the Arrhenius response for $n_\alpha$ units. The second term includes the probability factor when $n_T > n_\alpha$. The shift factor can then be written as: $a_T = a_A \times a_P$ where $a_A$ is due to the Arrhenius shift of $n_\alpha$ units and $a_P$ is due to the probability factor $e^{-n_{c(T)}}$ for additional $n_{c(T)}$ units imposed by limited free volume. Figure 7 compares shift factors obtained using the WLF equation and Equation 32 for poly(isobutyl methacrylate).

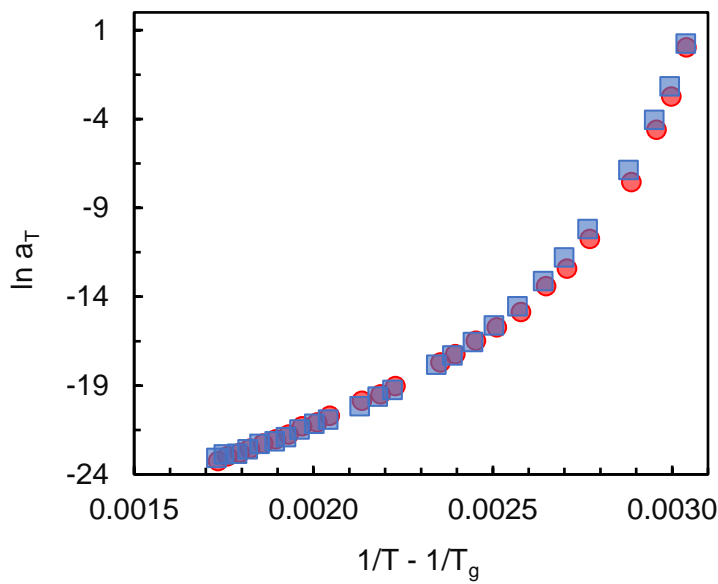

**Figure 7.** Example of shift factor calculations. The shift factor $a_T$ for poly(isobutyl methacrylate) calculated using WLF Equation 31 (red circles) and Equation 32 (blue squares). The empirical WLF fit constants were: $c_1 = 12.4$ and $c_2 = 57$ °C and $T_2 = T_g$. Values for $n_\alpha$, $E_a$ and $T_{SA}$ used for Equation 32 are given in Table 1.

**Crossover Temperatures.** While the WLF equation is mathematically equivalent to the VFT equation, it cannot be superimposed on Equations 27 or 32. The WLF and VFT equations are often unable to fit *both* the near-$T_g$ *and* the Arrhenius regions. Equation 27, in contrast, can describe the relaxation across wide ranges of temperatures. Equation 27 handily avoids the controversy of a temperature at which $\omega_\alpha$ would diverge (the Vogel temperature, $T_0$, section S4), close to the temperature $T_K$ at which the entropy of a glass would be equal to the entropy of its crystalline state.[78] This divergence is also avoided by many other descriptions of glass relaxation which use characteristic temperatures greater than $T_g$.[1] Such "crossover" temperatures define the point of departure from Arrhenius into supercooled behavior, much like $T_{SA}$. Stickel et al. presented way to emphasize the crossover region more sharply.[15] The resulting "Stickel temperature," $T_B$, is in the vicinity of $T_{SA}$ (e.g. $T_B = 243$ K for PPG[12]) or not very close (e.g. $T_B = 296$ K for OTP[12]). Martinez-Garcia et al. compared $T_B$ with the "critical temperature," $T_C$, from mode coupling theory and found $T_C$ to be in the vicinity of $T_B$.[12]

    **Cooperative Dynamics in Ion-Polymer Coupled Systems.** Many phenomena are coupled to structural dynamics.[79-80] For example, ion conductivity in polymers is usually shown to be coupled to the dynamics of the polymer host.[81-82] The dependence of conductivity on temperature for an example ion conducting polymer is modeled well by adapting Equation 27 using $n_\alpha = 2$ (Figure 8 and Supplemental Information Section S5).

$$\ln \sigma_T T = \ln \sigma_{T,Arr} T - exp\left(\frac{E_a}{2R}\left(\frac{1}{T} - \frac{1}{T_{SA}}\right)\right) \qquad [33]$$

We calculated $n_{Tg}$ to be 13. For a particular set of variables in Eq. 33, anything that increases the activation entropy will increase σ.

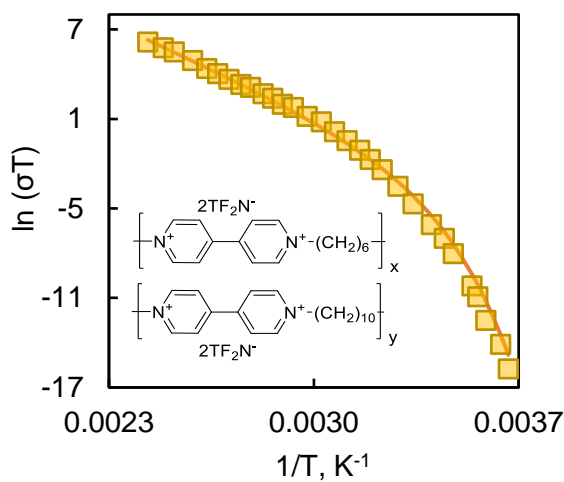

**Figure 8.** Conductivity plot of bis(trifluoromethylsulfonyl)imide,[83] $TF_2N^-$, in a single-ion polyviologen conductor (PV_$C_6$/$C_{10}$, structure given in inset). Solid line is Equation 33 using $n_\alpha = 2$, measured $E_a = 75.2$ kJ mol$^{-1}$, measured $T_{SA} = 317$ K.

This system is a single ion conductor, where the $TF_2N^-$ counterion is free to transport charge through the bulk of the material, whereas the positive charge resides on the polymer chain and is thus not free to diffuse.[83] The conductivity $\sigma T$ is proportional to the ion diffusion coefficient, $D_T$. The good agreement given in Figure 8 supports the idea that the transport of ions is coupled to the dynamics of the host polymer. Some ion transport below $T_g$ is claimed to be decoupled from polymer dynamics.[84] This may be true for the α-relaxation but there must still be some phonon-type of contribution from the matrix in which the ion is embedded. The "unphysically high prefactors" up to $10^{28}$ s$^{-1}$ for ion transport in polymer electrolytes recently highlighted by Gainaru et al.[84] must include a significant ΔS*.

**Cooperative Dynamics in Spin Systems.** Most of the temperature-controlled glass relaxation in the literature is fit to some reasonable extent by Equation 27. The fidelity of the data is somewhat degraded on transcribing literature plots and a few degrees of error in the temperature measurement near $T_{SA}$ would cause noticeable deviation (for an example see Figure S5B, Supplemental Information). Equation 27 broadly describes a "quenched" state where increasing numbers of cooperatively interacting units are obtained at lower temperatures with the emergence of heterogenous dynamic length scales. Spin glasses and ferromagnetic relaxors potentially fit this description. For example, Figure 9 shows relaxation dynamics for perovskite ferroelectrics PLZT ($Pb_{1-x}La_x(Zr_{1-y},Ti_y)_{1-x/4}O_3$) and PMN-PT[85] ($Pb(Mg_{1/3}Nb_{2/3})O_3$)-$PbTiO_3$ represented by Equation 27 with respective $n_\alpha$ values of 0.5 and 2.

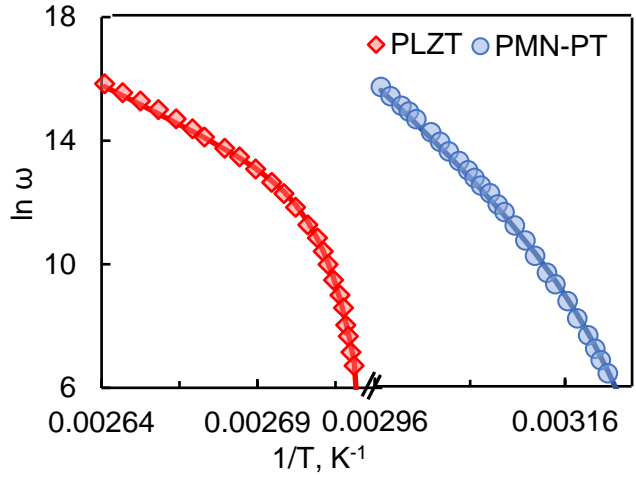

**Figure 9**. The frequency of ferroelectric relaxors versus 1/T. $Pb(Mg_{1/3}Nb_{2/3})O_3 - PbTiO_3$ (PMN-PT, ○): $n_\alpha = 2$, experimental $E_a = 229$ kJ mol$^{-1}$, measured $T_{SA} = 319$ K; and $Pb_{(1-x)}La_x(Zr_{1-y}Ti_y)_{1-x/4}O_3$ (PLZT, ◊): $n_\alpha = 0.5$, experimental $E_a = 422$ kJ mol$^{-1}$, measured $T_{SA} = 369$ K . Points are experimental data from ref [85]; solid line from Equation 27.

In the case of electric (or magnetic) dipoles we interpret an $n_\alpha$ of 0.5 to represent one dipole switching direction from +1 to -1 (rather than +1 to 0).

## Conclusions

We have presented an equation based on measurable parameters that allows quantitative description and classification of glassy dynamics according to the minimum number of units that must rearrange simultaneously. The motion of units has been broken down into those occurring simultaneously (concerted/correlated) and those occurring within interval $\tau_1$. $n_\alpha$ represents the minimum cluster size of rearranging units, observed at high temperatures. These units escape their cages simultaneously with frequency $\omega_\alpha$, but only if there is enough free volume to do so. Otherwise, one observes a correlated rattle at $\omega_{JG}$, the JG β-relaxation. $\omega_{JG}$ accounts for the Arrhenius slope of $\omega_\alpha$ at high temperature. Additional spatiotemporal restrictions are encountered as the temperature drops, limiting the free volume. A probability term incorporates dynamic heterogeneity and deviation from Arrhenius due to a growing length scale of cooperativity as the glass cools.

The only freely-adjustable fitting term is $n_\alpha$, which is an integer between 1 and 4. A mixture of kinetics and thermodynamics is inherent to Equation 12. The $E_a$ term refers to the steady-state (equilibrium) Boltzmann distribution of the number of $n_\alpha$ units with energy sufficient to simultaneously escape their cages. The $n_{c(T)}$ term is a result of additional kinetic restrictions. Though our approach can be classified as a free volume argument, the actual free volume is not needed. Neither the free volume at $T_g$, nor $T_g$ itself is specified unless $n_{T_g}$ is estimated for a class of materials, such as polymers, for Equation 30. In fact, the most important temperature regime is around $T_{SA}$ because it provides $E_a$ and $T_{SA}$. With the Arrhenius slope, $T_{SA}$, and $n_\alpha$, the entire supercooled regime between $T_{SA}$ and $T_g$ is predicted with reasonable accuracy.

The results are entirely consistent with work attributing certain β-relaxations to caged precursors,[8, 34, 86] which are unleashed above $T_g$ as groups of $n_\alpha$ units assisted by cooperative motions among $n_\alpha + n_{c(T)}$ units. Extrapolations of $\omega_\alpha$ to the liquid state (i.e. $T > T_{SA}$) provide deep implications on how liquids flow. For example, substances with $n_\alpha > 2$ move in clusters of $n_\alpha$, not as individual units. Knowing $n_\alpha$ should improve molecular simulations and vice versa. The persistence length exchange mechanism suggested for polymers (Figure 3), if it holds up under scrutiny from molecular dynamics simulations, may be universal for polymers. The $n_\alpha = 2$ found for ionic liquids may also be universal, and may represent the place exchange of two neighboring like-charged ions to preserve the local electrostatic field.

While Equation 27 brings much-needed simplicity[66] to a quantitative relationship for glassy dynamics *versus* temperature, many challenges remain, including developing a more sophisticated picture of both the nature of a "unit", which is not necessarily spherical, and its interaction environment.

Some assumptions have been made which are not strictly valid. For example, whatever high frequency mode actually represents $A_1$ shows slight temperature dependence.[56] The activation barrier $E_1$ is also likely not constant over the supercooled region, though there may be some fortuitous cancelling of effects. A changing $E_1$ translates to a nonlinear Arrhenius response, which is more likely to occur with dynamics that are studied over a wide temperature range, such as those of OTP.

**Author Information**

*Corresponding Author*

**J.B. Schlenoff** - Department of Chemistry and Biochemistry, The Florida State University, Tallahassee, Florida 32306-4390, United States. Email: jschlenoff@fsu.edu

*Authors*

**K. Akkaoui** - Department of Chemistry and Biochemistry, The Florida State University, Tallahassee, Florida 32306-4390, United States

**Acknowledgments**

This work was supported by the National Science Foundation (grants DMR-1809304 and DMR-2103703).

**Conflict of Interest**

The authors have no conflicts to disclose.

**Supplementary Information**

Relaxations diagrams for three alcohols; individual relaxation plots of the glass formers transcribed from the literature, alongside $E_a$, $A_\alpha$ and $T_{SA}$; detailed derivation of Equation 27; comparison with VFT and WLF equations; Equation 27 adapted for ionic conductivity.

**Supplemental Information**

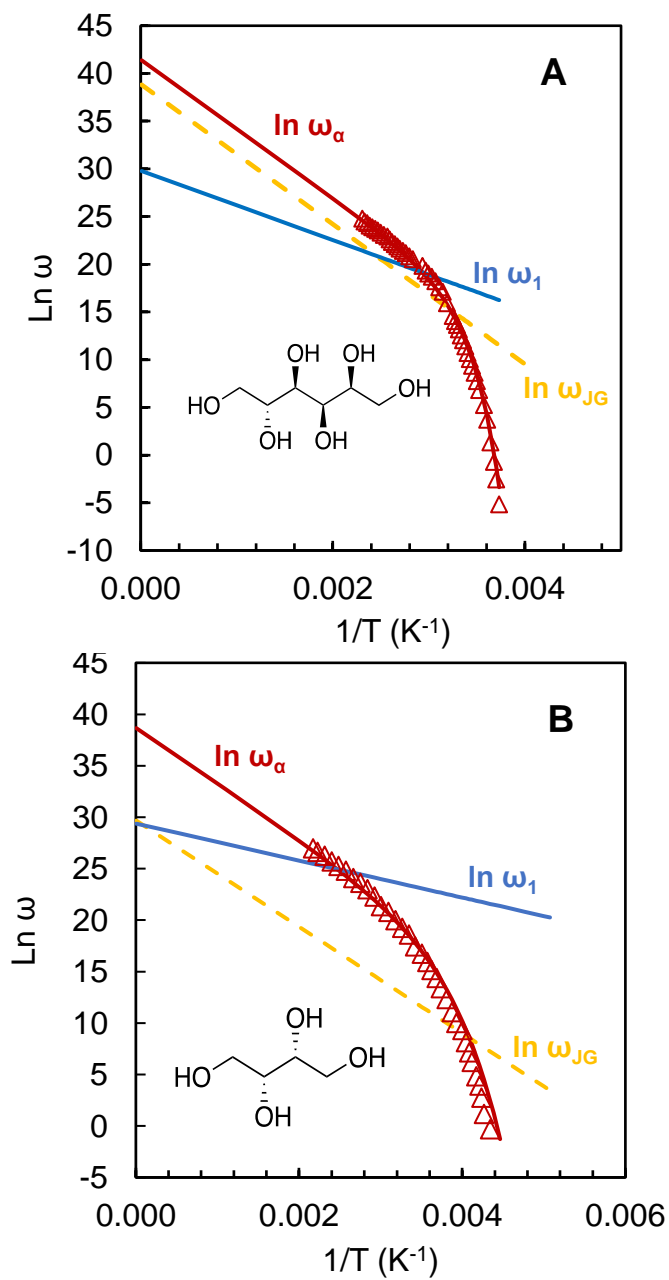

**Figure S1.** Relaxations in two alcohols: **A**, sorbitol; **B**, threitol. Rates $\omega_\alpha$ and $\omega_{\beta JG}$ and $\omega_1$. $\beta_{JG}$ relaxation (s$^{-1}$) for sorbitol (n$_\alpha$ = 2) and threitol (n$_\alpha$ = 3) as a function of 1/T (dashed line). $\omega_1$ (slope = E$_1$/R). Respective $A_\alpha$ = 1 x 10$^{18}$ s$^{-1}$ and 6.3 x 10$^{16}$ s$^{-1}$; E$_a$ = 60.5 kJ mol$^{-1}$ and 44.8 kJ mol$^{-1}$. T$_{SA}$ = 340 and 342 K for sorbitol and threitol, respectively. A$_1$ = 8.8 x 10$^{12}$ and 5.8 x 10$^{12}$.[88-89] The $\beta_{JG}$ relaxations are from reference 46: A$_{JG}$ = 7.7 x 10$^{16}$ s$^{-1}$ and 7.9 x 10$^{12}$ s$^{-1}$, E$_{a,JG}$ = 61 kJ mol$^{-1}$ and 43 kJ mol$^{-1}$.

### Section S1: Probability for unit move

A Poisson distribution describes the probability that a given number of random events will occur in a fixed interval of time, and is given by a probability mass function, p(k), with k the number of occurrences: $p(k) = \frac{\lambda^k}{k!}e^{-\lambda}$. The probability, p(k = 1), of a specific unit (k = 1) moving in its cage in $\tau_1$ seconds, with an average rate $\lambda$ = 1 move per $\tau_1$ seconds, is $p(k) = \frac{1^1}{1!}\left(\frac{1}{e}\right)$.

### Section S2: Free volume

The purpose of $n_{c(T)}$ unit movement is to concentrate free volume around $n_\alpha$ units. One can estimate the outcome of this for polymers ($n_\alpha$ = 2). Suppose the minimum local fractional free volume required for 2 units to slide past each other (see Figure 4D) is $v_{f\alpha}$. If the average free volume of a unit at a particular temperature is $v_{fl(T)}$, the total localized free volume is $n_T v_{fl(T)}$. For polymers, above $T_{SA}$, $2v_{fl(T)} > v_{f\alpha}$. Below $T_{SA}$, $2v_{fl(T)} < v_{f\alpha}$, so $n_{c(T)}$ units must contribute additional free volume.

### Section S3:

Equation 22 is reproduced here.

$$\frac{d \ln \omega_\alpha}{d\, T^{-1}} - \frac{E_1}{R} \ln \omega_\alpha + \frac{E_1}{R}\left(n_\alpha + \ln A_\alpha - \frac{n_\alpha E_1}{RT}\right) = 0 \qquad [S1]$$

Equation S1 is a first order non homogenous differential equation of the form: y' +ay +bx +c =0 where $y = \ln \omega_\alpha$, $x = T^{-1}$ and $a = -\frac{E_1}{R}$, $b = -\frac{n_\alpha E_1^2}{R^2}$, $c = \frac{E_1}{R}(n_\alpha + \ln A_\alpha)$. The solution is given by $y = \frac{b}{a^2} - \frac{b}{a}x - \frac{c}{a} + c_1 e^{\frac{E_1}{RT}}$

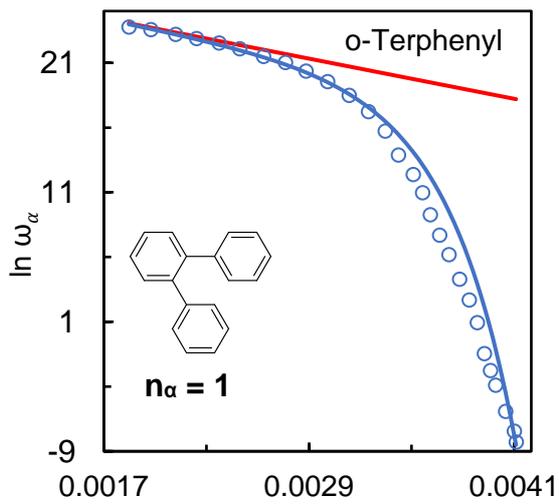
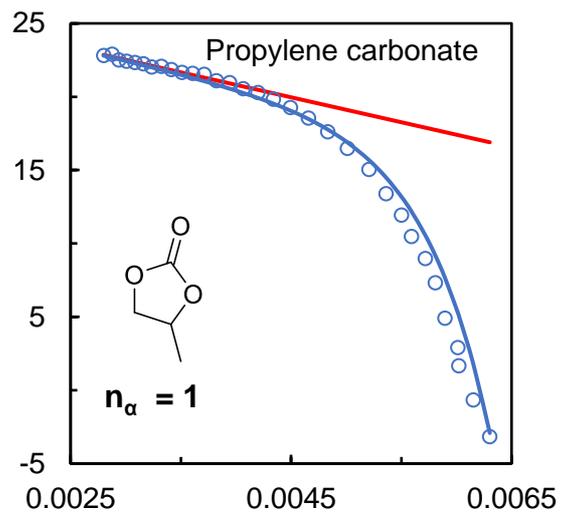
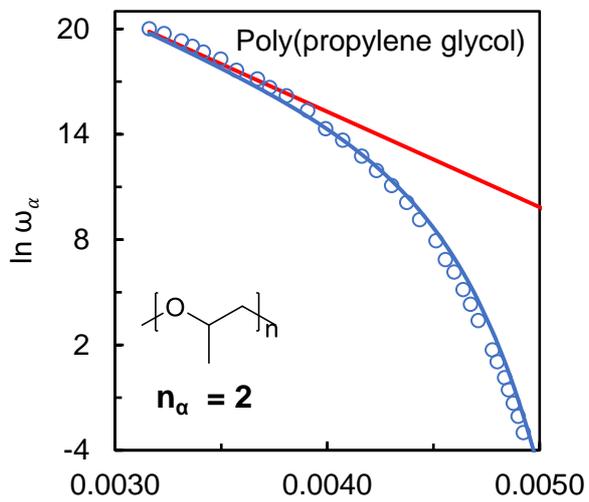
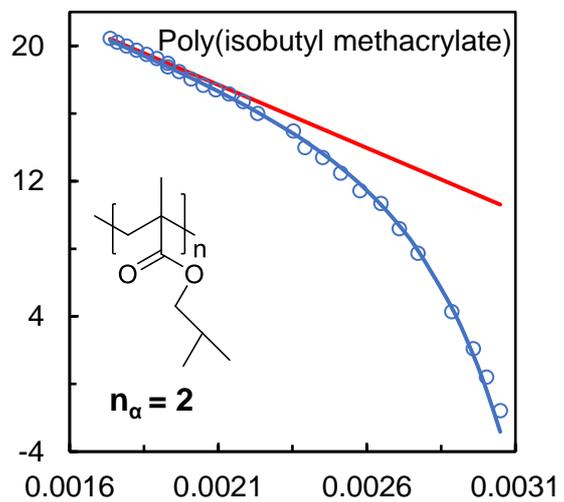
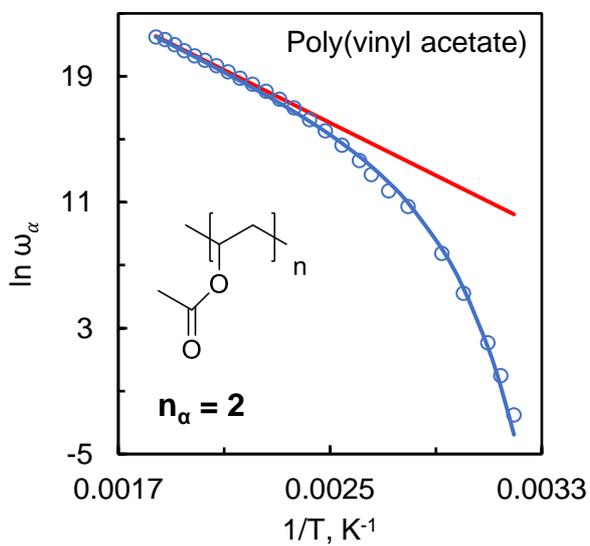
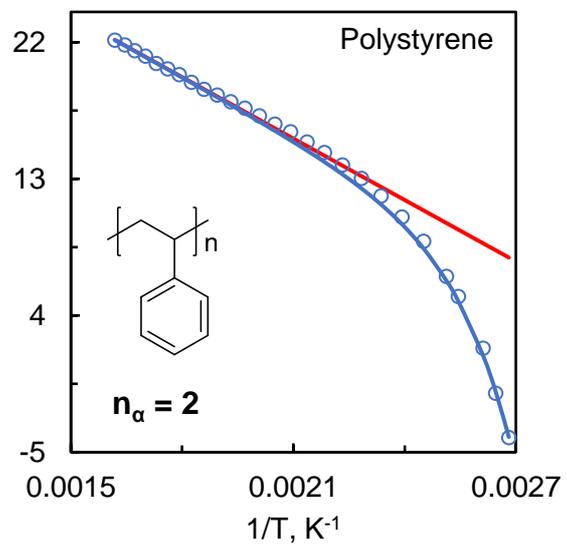

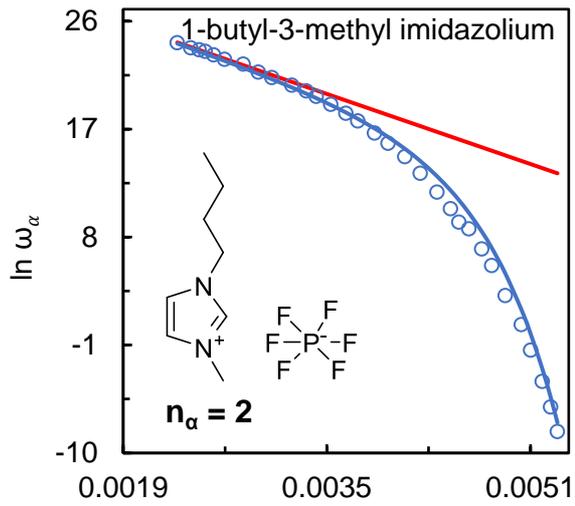
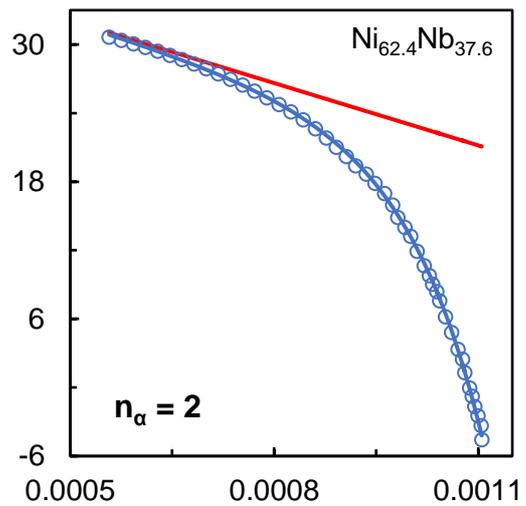
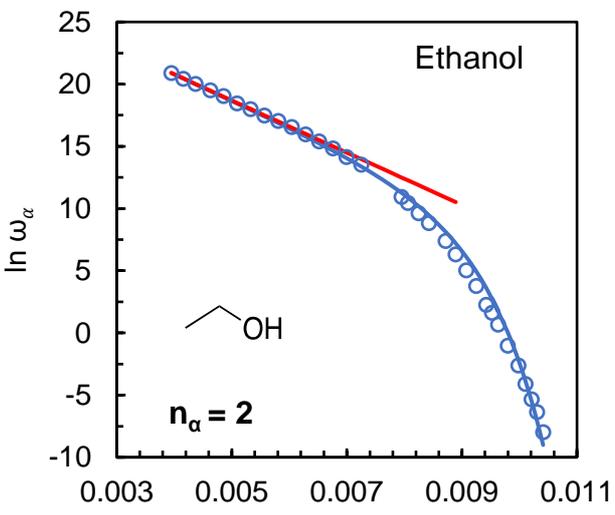
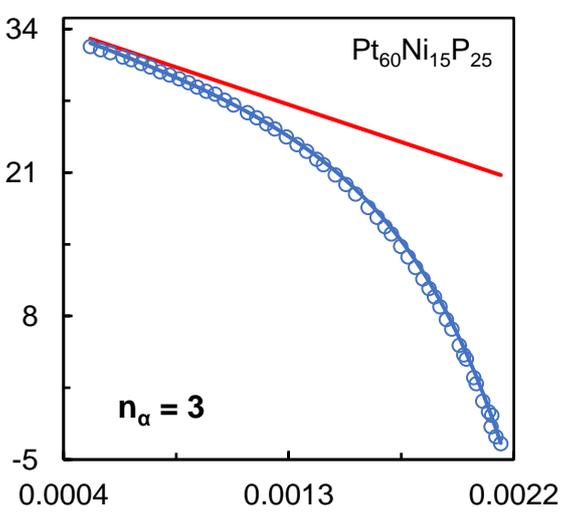
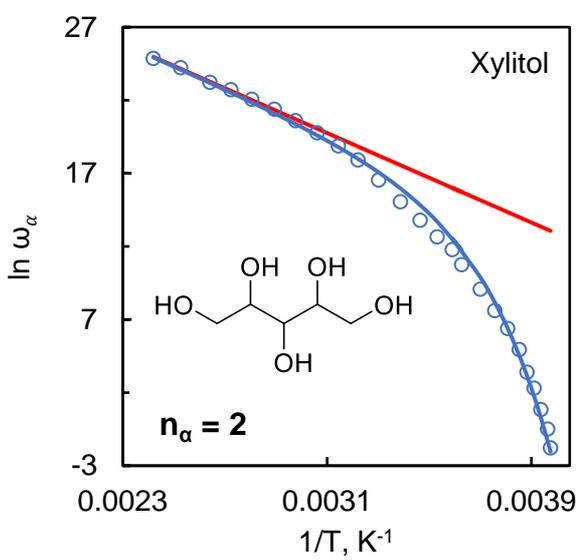
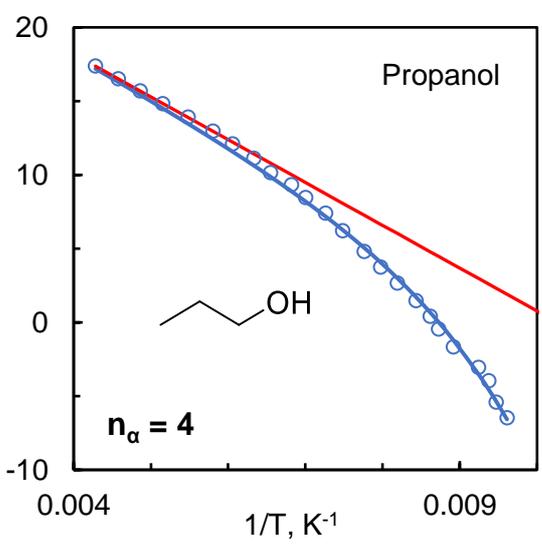

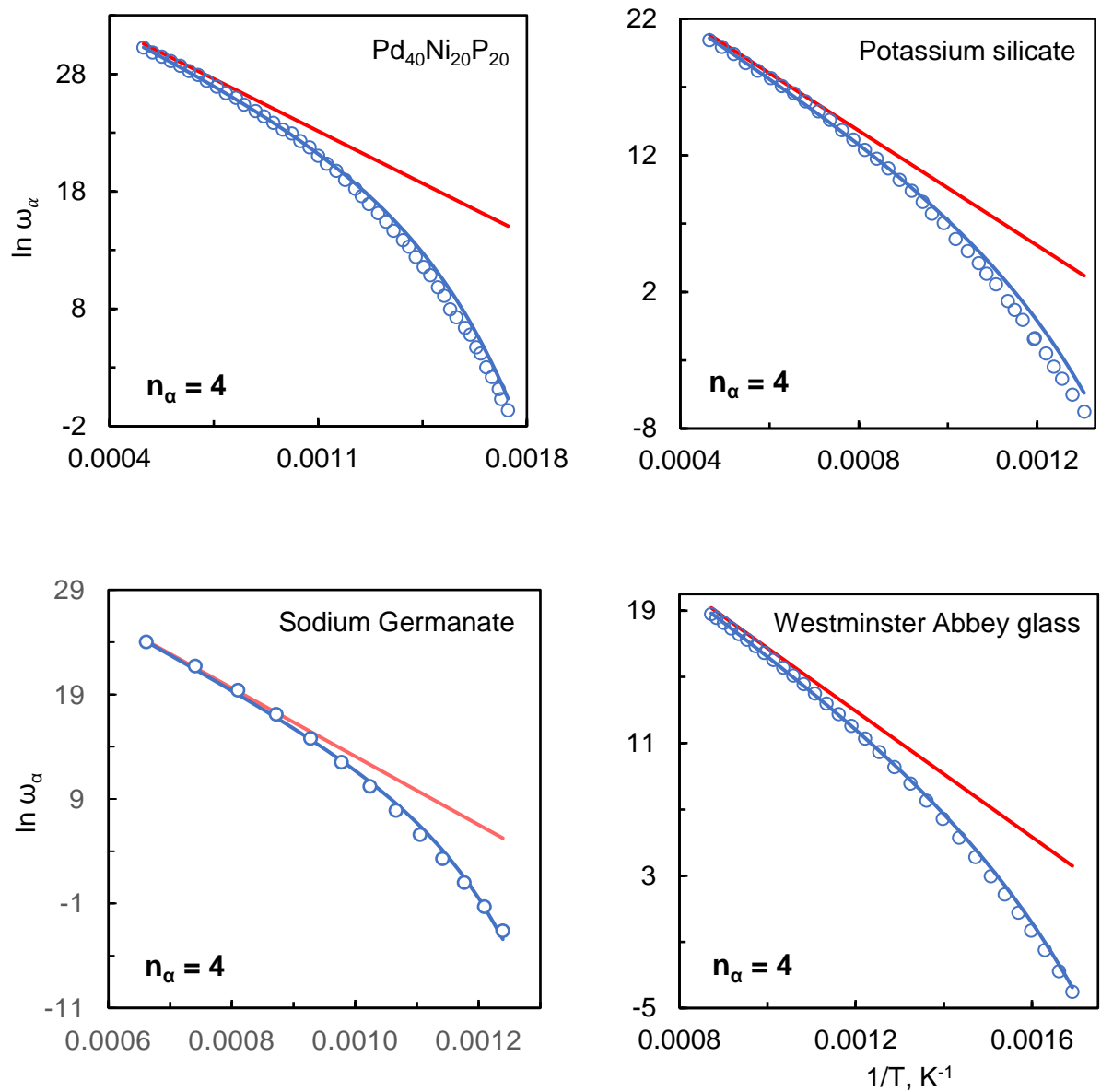

**Figure S2**. Relaxation in glass-forming liquids. The structural relaxation rate in 16 glasses given by experimental data transcribed from the literature (open circles), comparison to Equation 27 (blue curve) using the parameters given in Table 1, and the Arrhenius frequency (red line). Insets show the $n_\alpha$ class and the structure (when applicable).

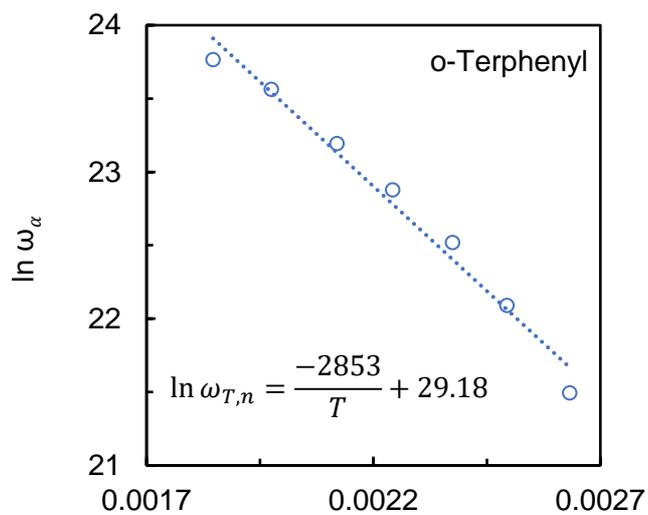
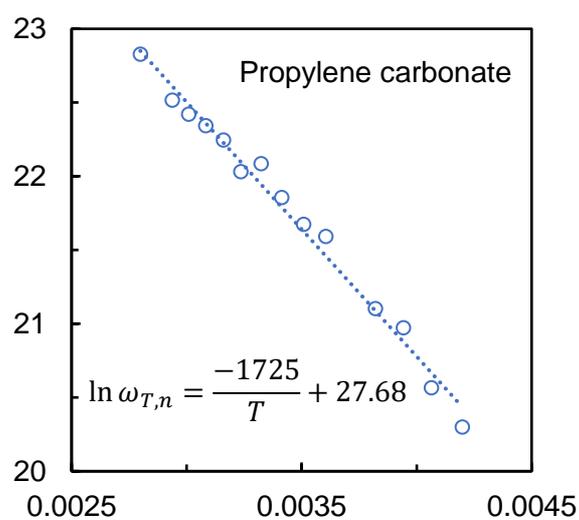
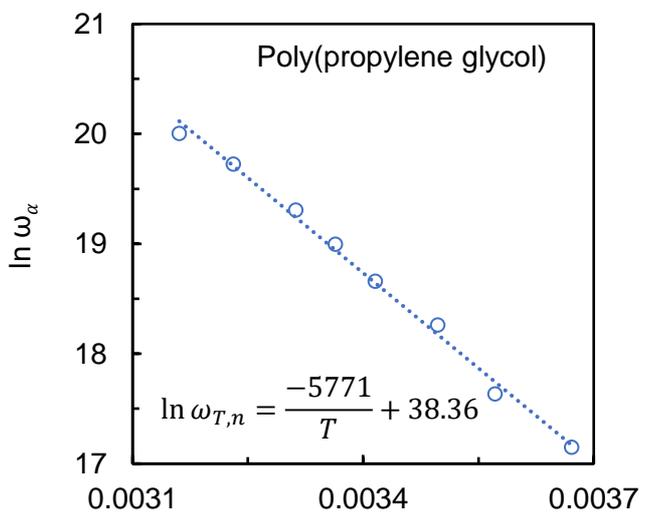
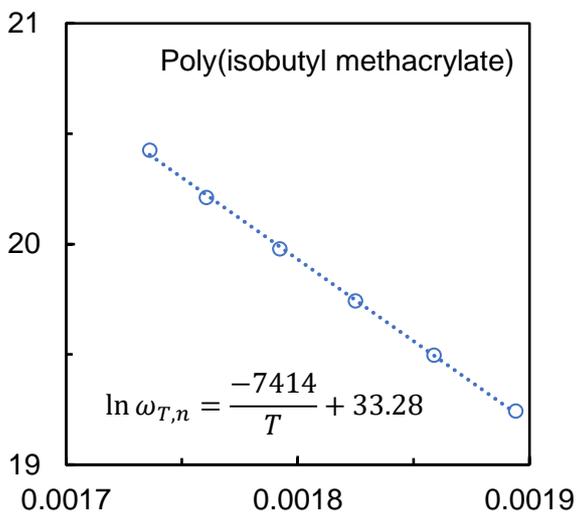
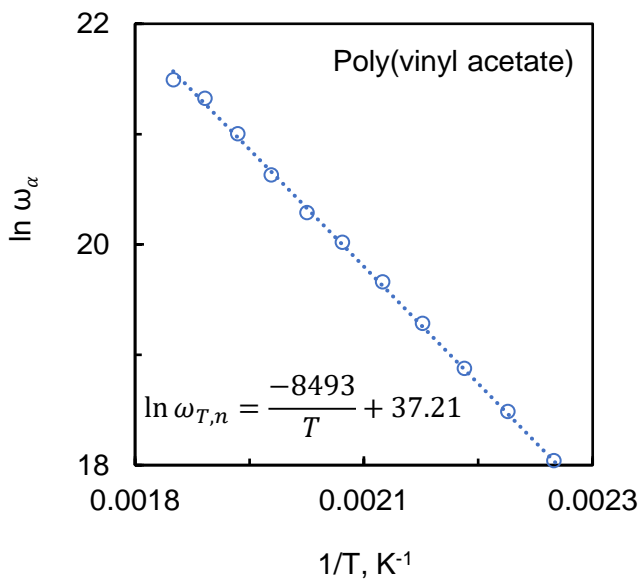
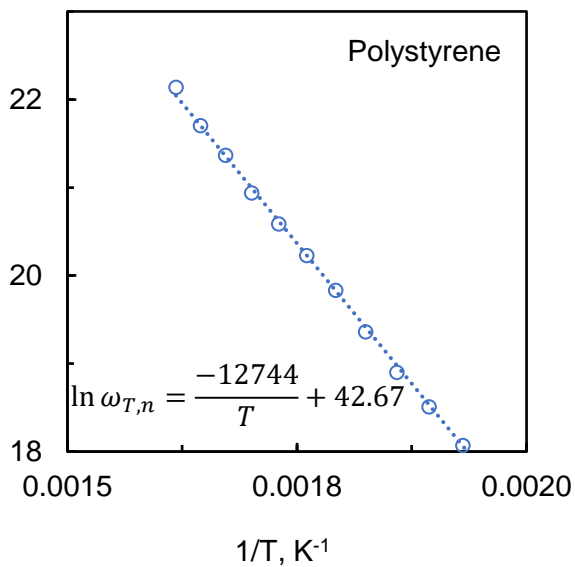

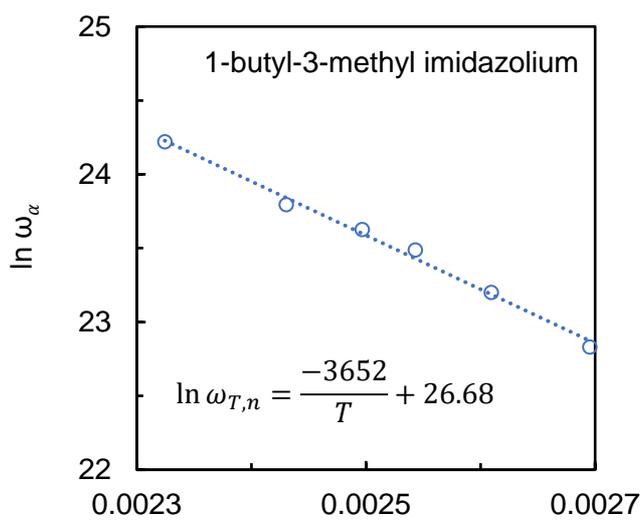
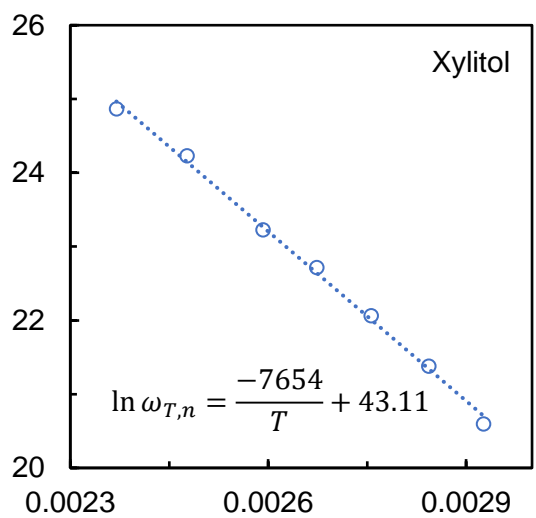
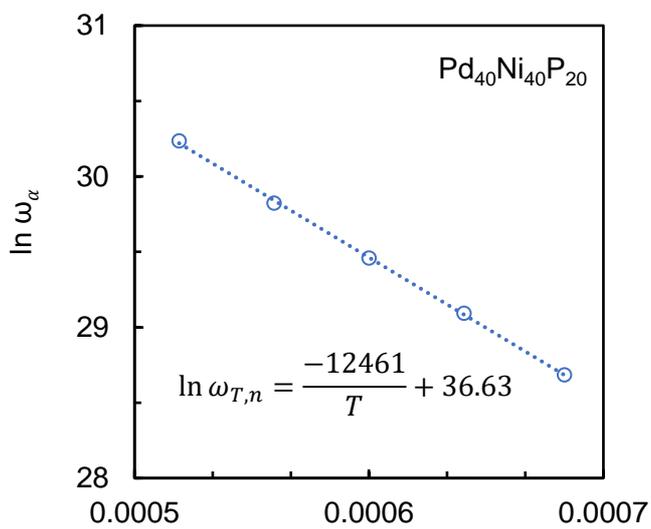
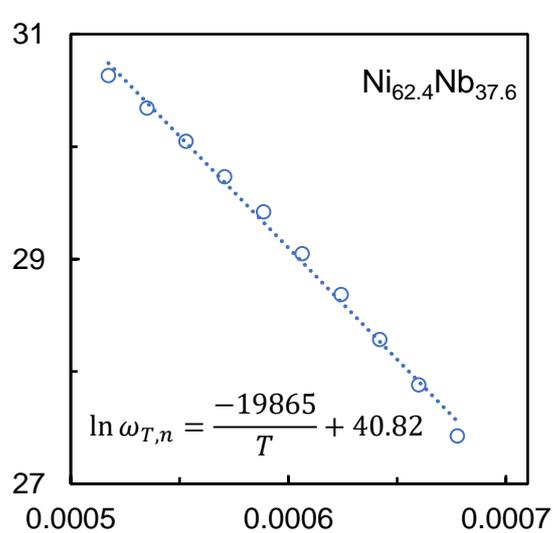
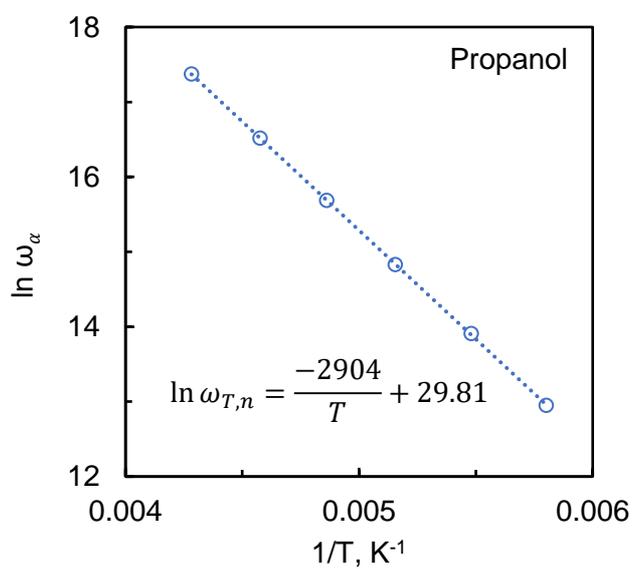
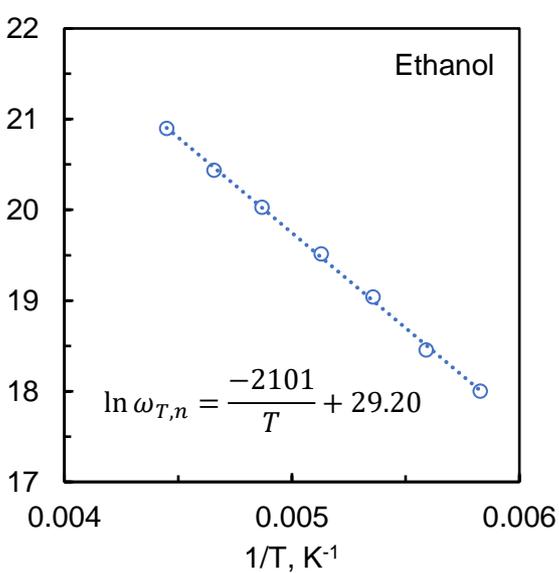

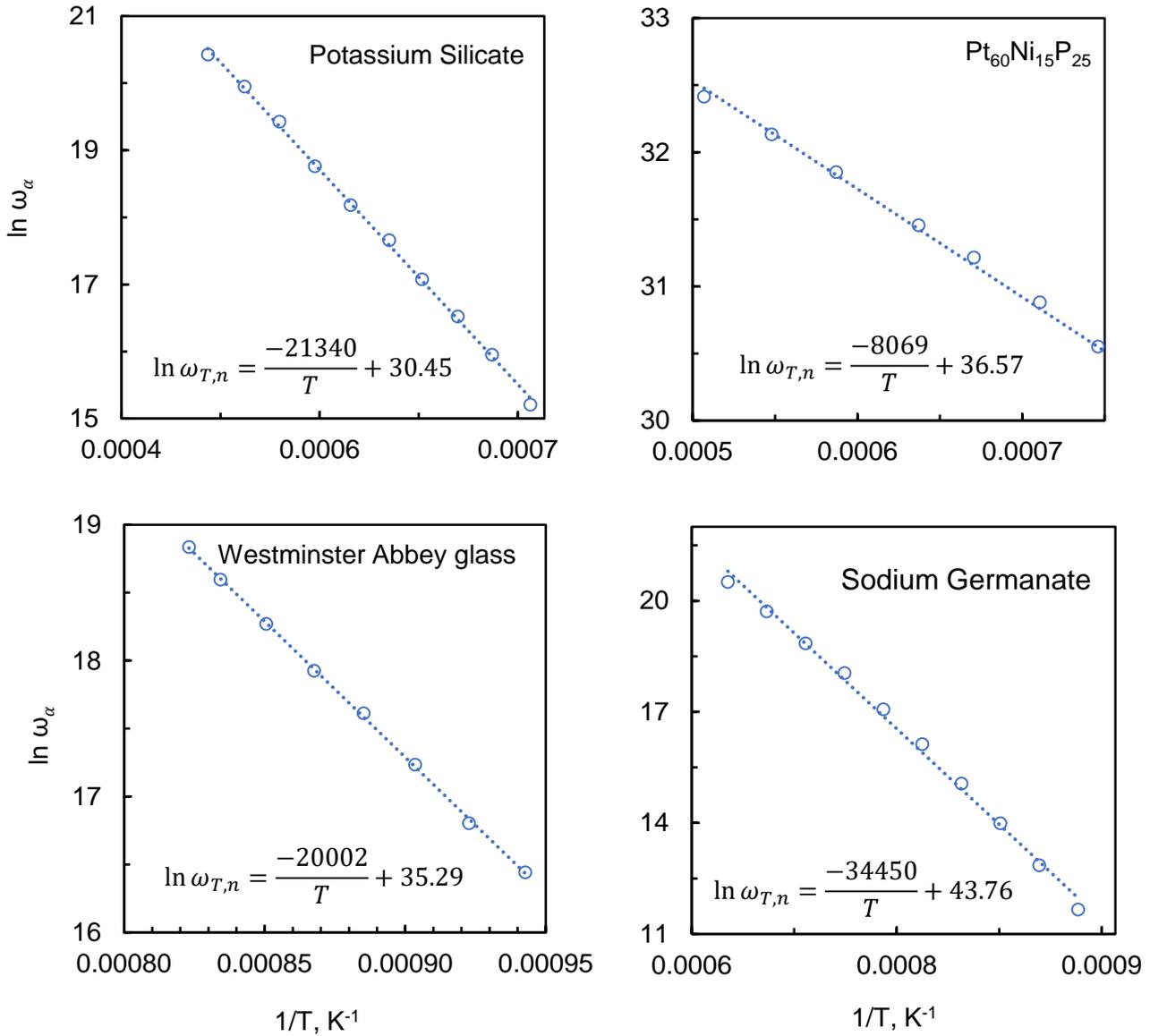

**Figure S3.** Estimating $E_a$ and $A_\alpha$ from high T data. Relaxation frequencies, $\omega_\alpha$ (s$^{-1}$) given in the Arrhenius region (T > $T_{SA}$) for the glass formers analyzed *versus* 1/T. The blue circles are data transcribed from the literature. Blue dotted line is a linear fit of *ln* frequency *vs.* 1/T (fit equation shown). The molar activation energies, $E_a$, are calculated by multiplying slopes by the gas constant (R = 8.314 J K$^{-1}$ mol$^{-1}$). The intercept gives the preexponential factor *ln* $A_\alpha$.

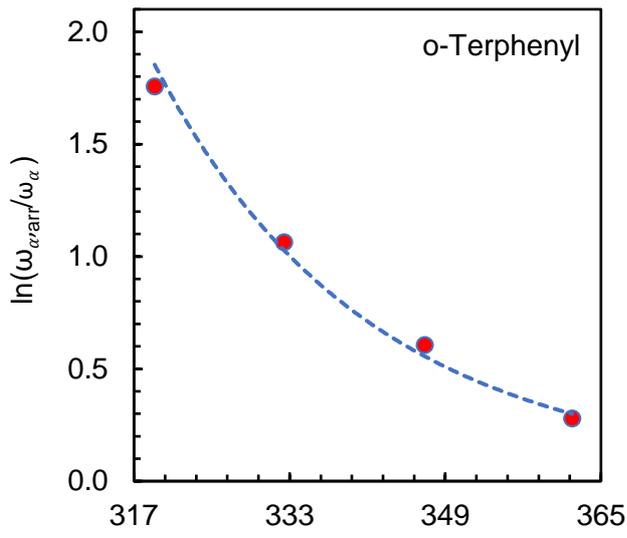
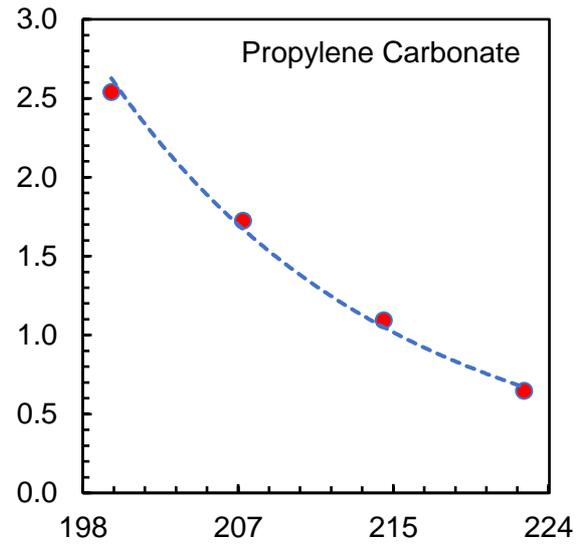
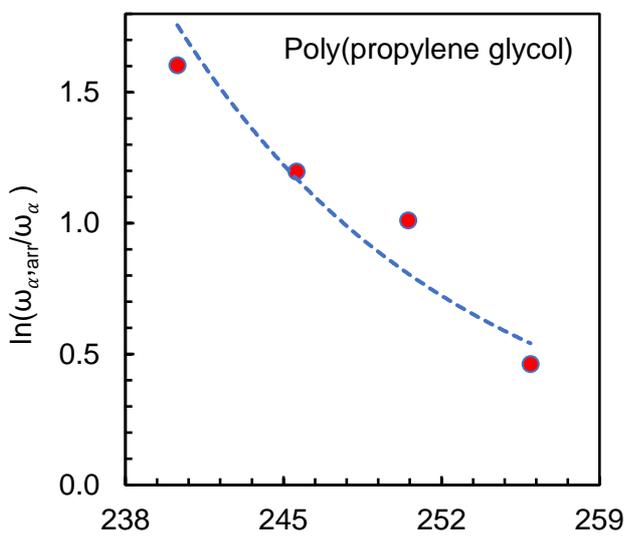
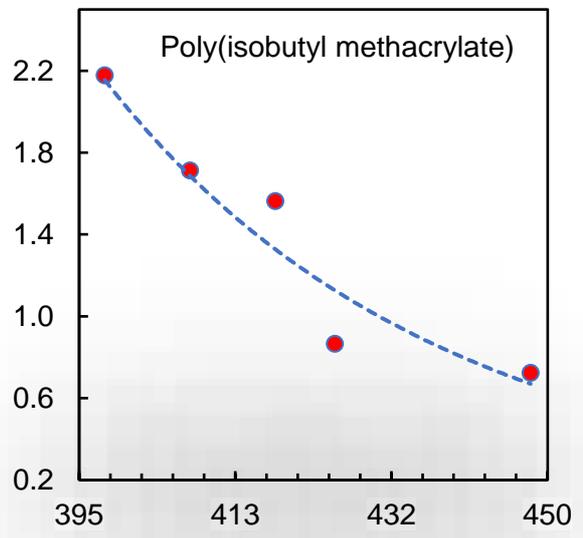
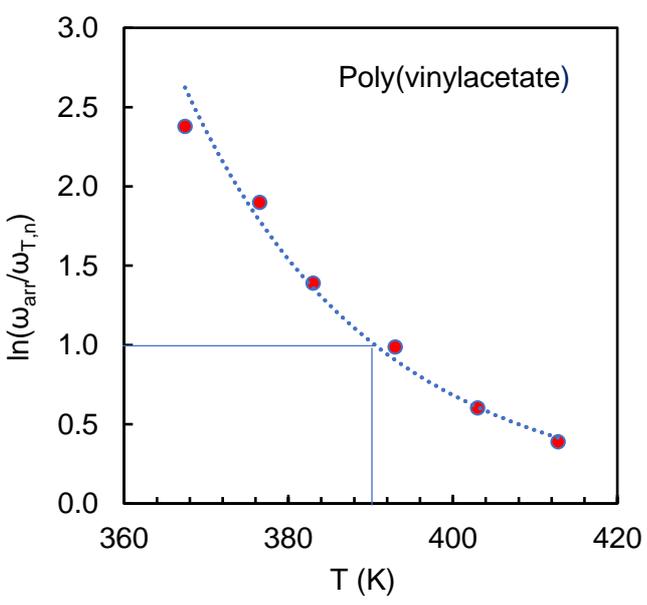
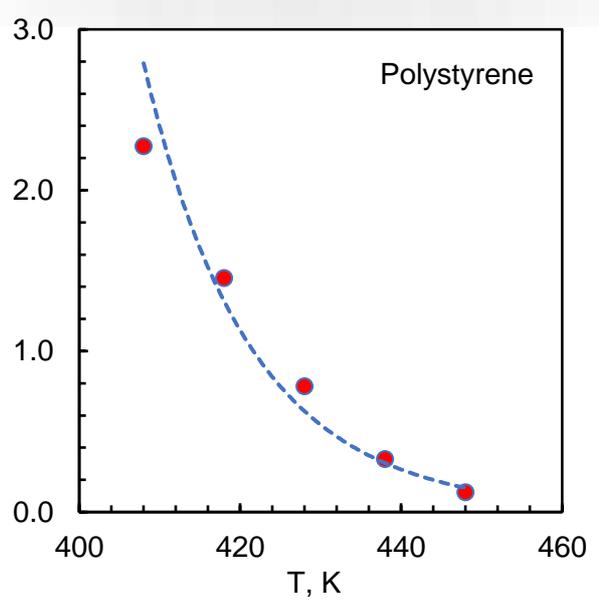

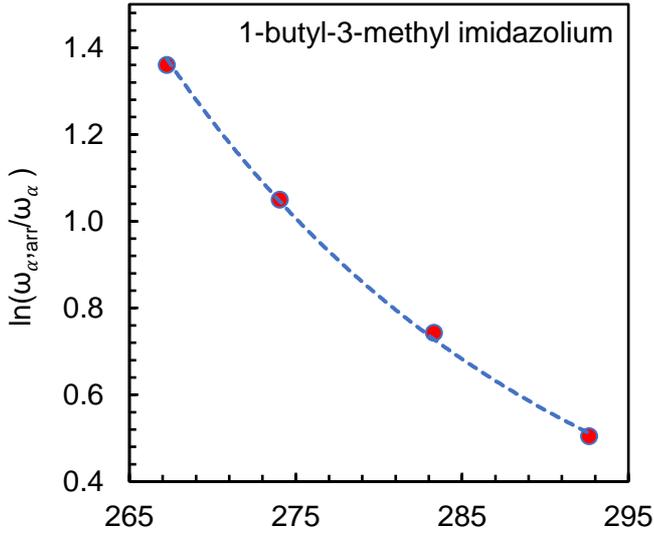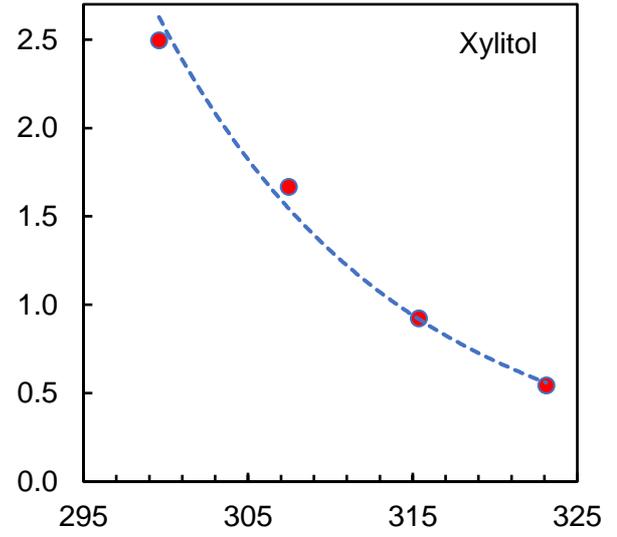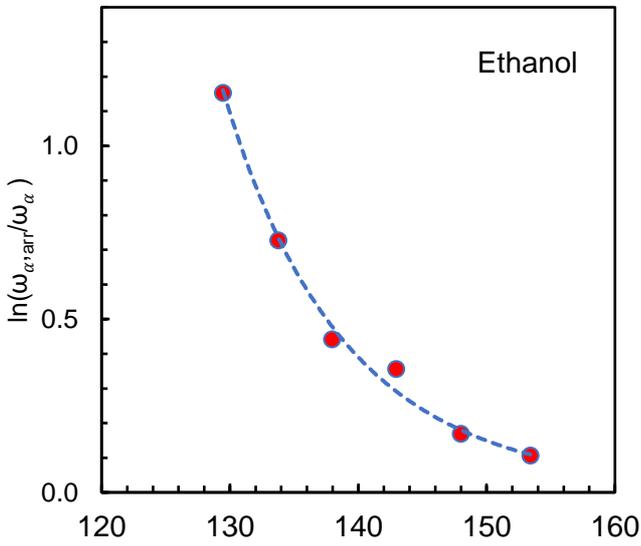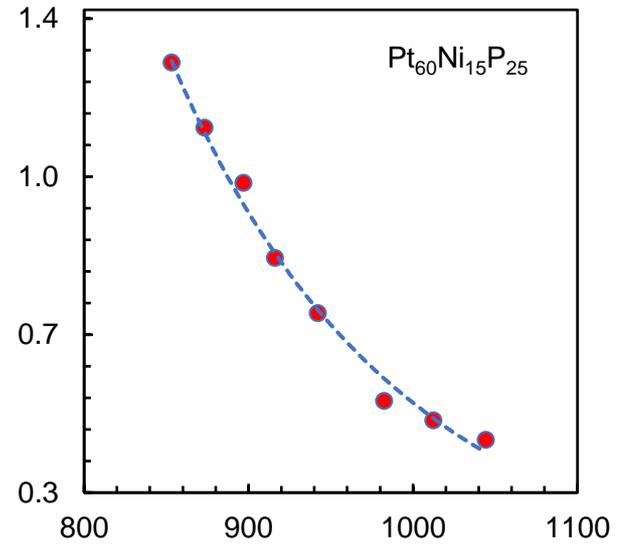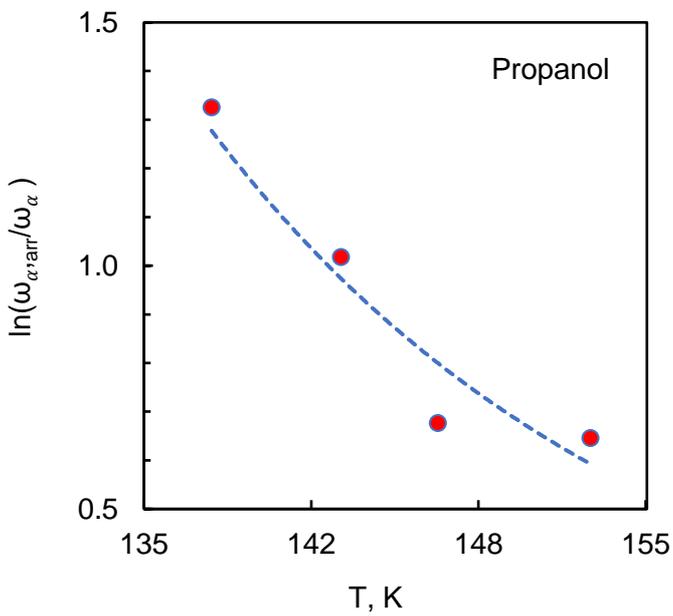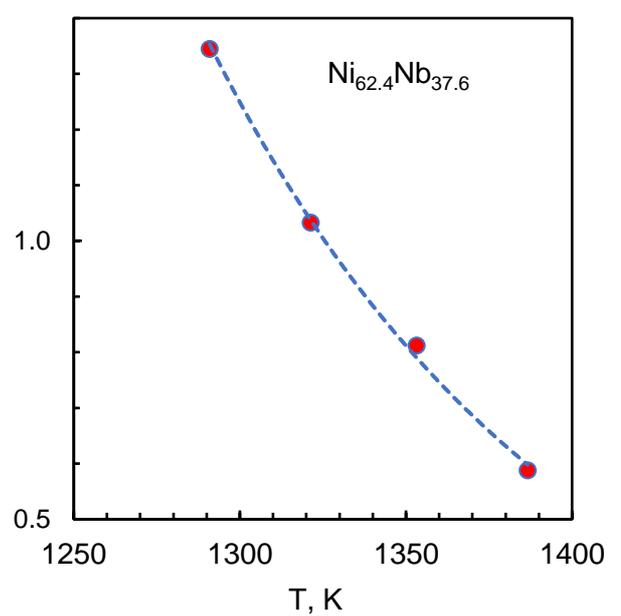

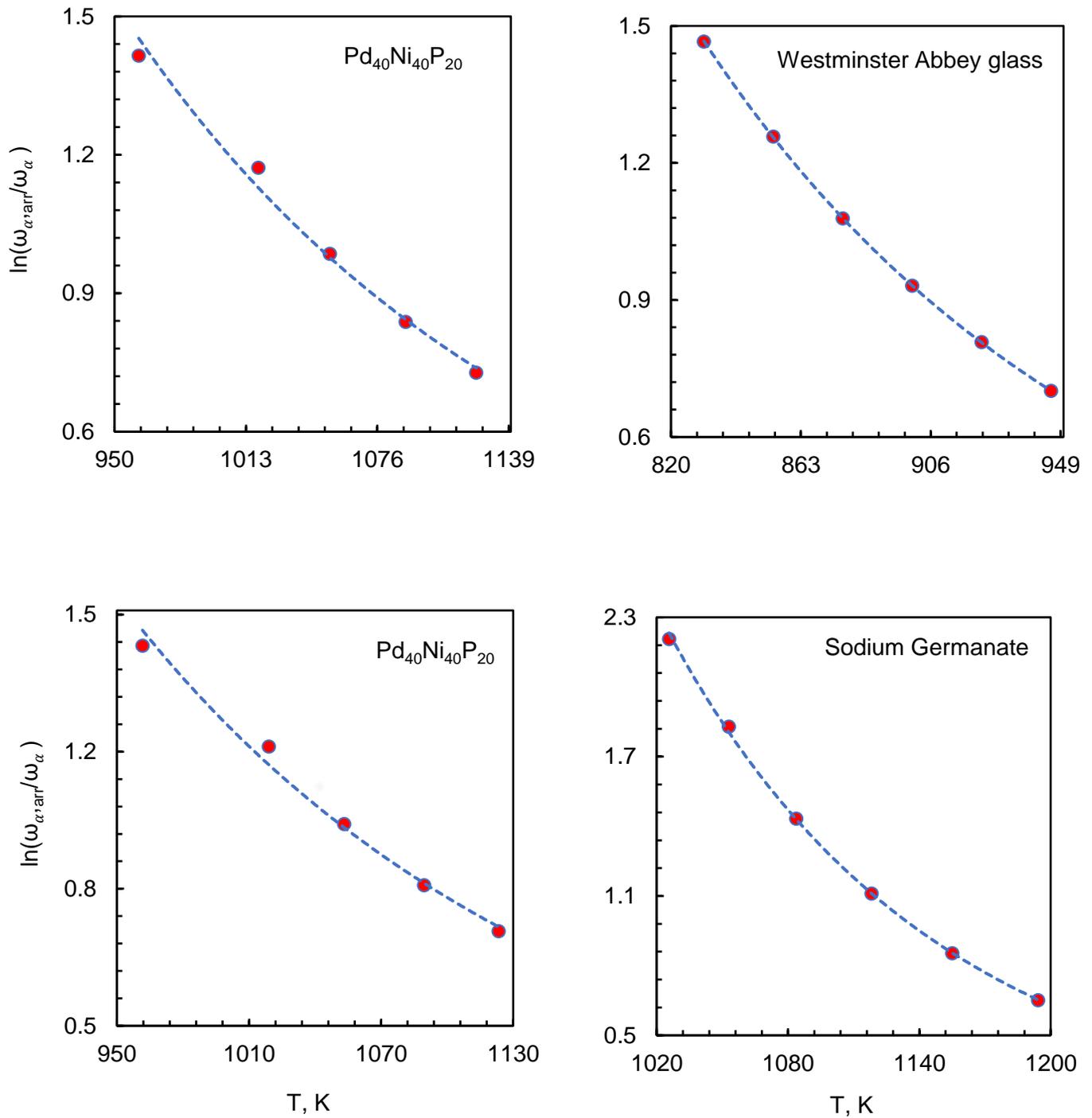

**Figure S4.** Locating $T_{SA}$. Plots of $\ln(\omega_{\alpha,arr}/\omega_\alpha)$ as a function of temperature allow the estimation of $T_{SA}$ using experimental data (at $T_{SA}$, $\ln(\omega_{\alpha,arr}/\omega_\alpha) = 1.0$). The circles are experimental data within an appropriate range for $T_{SA}$ estimation. The dashed lines are guides to the eye.

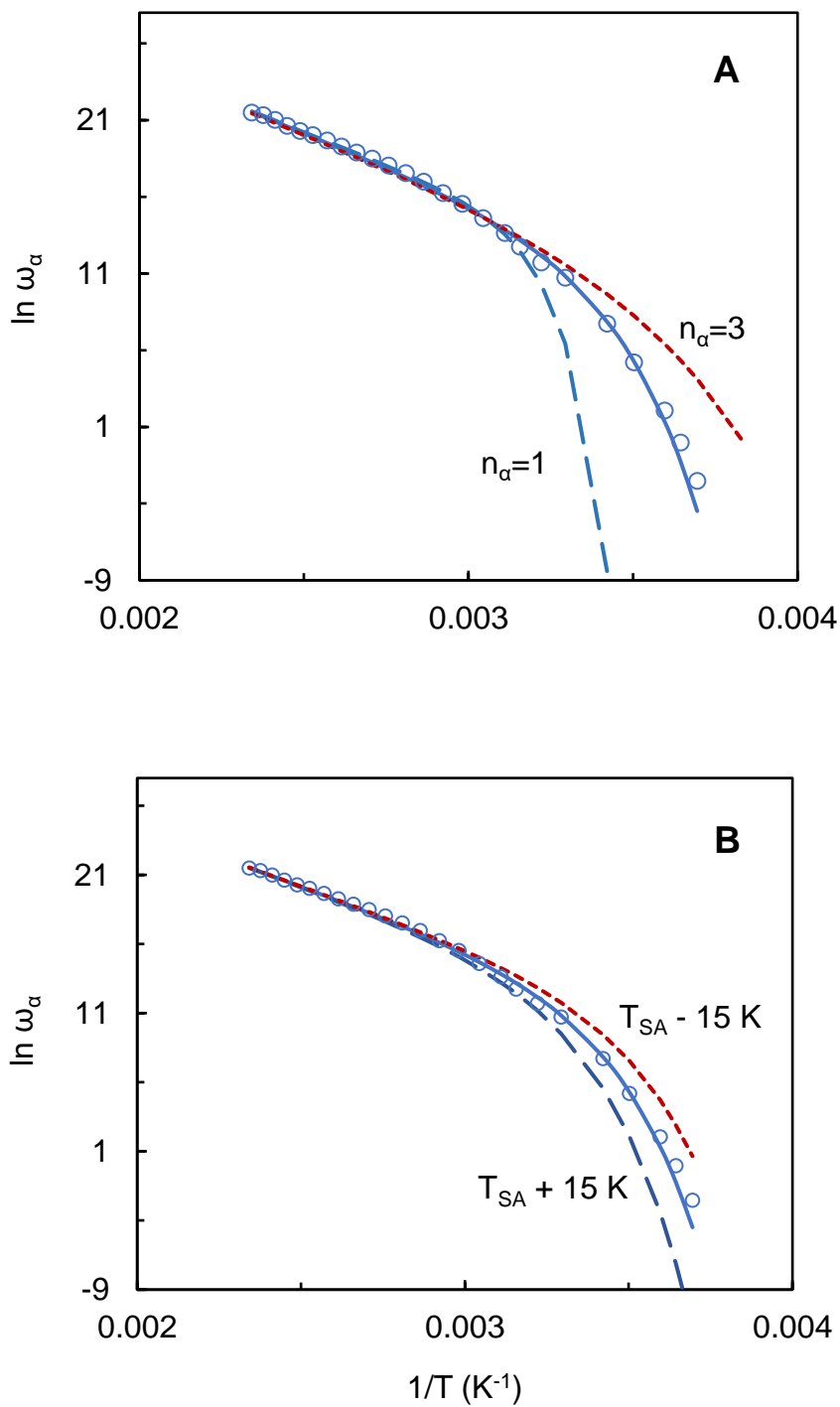

**Figure S5**. (**A**) Varying $n_\alpha$. Varying $n_\alpha$ in poly(vinylacetate) ($n_\alpha = 2$). The circles are experimental data taken from the literature (see Table 1 for the reference). The solid curve is the fit for $n_\alpha = 2$. The red short-dashed curve is the fit using $n_\alpha = 3$ and the blue long-dash curve is the fit using $n_\alpha = 1$. (**B**) Varying $T_{SA}$. The solid blue curve uses $T_{SA} = 290$ K as read from Figure S4. Red short-dash curve is $T_{SA} = 305$ K. Blue long-dash curve is $T_{SA} = 275$ K.

## Section S4: Shift factor $a_T$

The Williams-Landel-Ferry (WLF) equation defines a shift factor, $a_T$, used for time-temperature superposition, usually on polymers. It gives a ratio of the rate, $\omega_{\alpha,T}$ at two different temperatures $T_1$ and $T_2$

$$a_T = \frac{\omega_{\alpha,T_1}}{\omega_{\alpha,T_2}} \qquad [S2]$$

One temperature is usually selected as a reference (often $T_g$). The WLF equation gives an empirical relationship between $a_T$, $T_1$ and $T_2$ and two empirical fit constants $c_1$ and $c_2$ adjusted to fit the data

$$\log(a_T) = \frac{-c_1(T_1 - T_2)}{c_2 + T_1 - T_2} \qquad [S3]$$

The mathematically equivalent VFT equation usually fits relaxation response in the range of temperature $T_g$ to about $T_g + 100$ K

$$\omega_{T,VFT} = \omega_{0,VFT}\, e^{\frac{-DT_o}{T-T_o}} \qquad [S4]$$

Where $\omega_{0,VFT}$, D and $T_o$ are empirical fit parameters. $T_o$ is known as the Vogel temperature and is usually about 50 K below $T_g$. The equivalent form for viscosity, $\eta$, is

$$\eta_{T,VFT} = \eta_{o,VFT}\, e^{\frac{DT_o}{T-T_o}} \qquad [S5]$$

Combining Equations 27 and S2 gives a form of the shift factor between $T_1$ and $T_2$ using $E_a$, $T_{SA}$, and $n_\alpha$.

$$\ln a_T = \frac{E_a}{R}\left(\frac{1}{T_2} - \frac{1}{T_1}\right) + e^{\frac{-E_a}{n_\alpha R T_{SA}}}\left(e^{\frac{E_a}{n_\alpha R T_2}} - e^{\frac{E_a}{n_\alpha R T_1}}\right) \qquad [S6]$$

The WLF equation does not map directly onto Equation S6 but a new constant $c$ may be defined as follows

$$c = \frac{E_a}{n_\alpha R} \qquad [S7]$$

Therefore,

$$\ln a_T = \frac{E_a}{R}\left(\frac{1}{T_1} - \frac{1}{T_2}\right) - e^{-\frac{C}{T_{SA}}}\left(e^{\frac{C}{T_2}} - e^{\frac{C}{T_1}}\right) \quad [S8]$$

## Section S5: Diffusion and ion conductivity

Equation 27 adapted for the dependence of ion conductivity on temperature in polymers ($n_\alpha = 2$) reads

$$\ln D_T = \ln D_{Arr} - exp\left(\frac{E_a}{2R}\left(\frac{1}{T} - \frac{1}{T_{SA}}\right)\right) \quad [S9]$$

It follows Arrhenius behavior at sufficiently high temperatures

$$D_{T,Arr} = D_0 e^{-E_a/RT} \quad [S10]$$

The Nernst-Einstein equation relates conductivity $\sigma_T$ to $D_T$ in a single-ion conductor:

$$\sigma_T = \frac{q^2 C D_T N_A}{k_B T} \quad [S11]$$

where $N_A$ is Avogadro's number, $C$ is the ion concentration (moles m$^{-3}$), $k_B$ is Boltzmann's constant and $q$ is the charge of the ion. The conductivity would read

$$\ln \sigma_T T = \ln \sigma_{T,Arr} T - exp\left(\frac{E_a}{2R}\left(\frac{1}{T} - \frac{1}{T_{SA}}\right)\right). \quad [S12]$$

For $T > T_{SA}$, assuming $C$ does not change significantly with temperature

$$\sigma_T T = A_\sigma e^{-E_a/RT} \quad [S13]$$

where $A_\sigma = \frac{q^2 C D_0 N_A}{k_B}$.